\documentclass[useAMS,usenatbib,babel,times]{mn2e}
\usepackage{multirow,balance,verbatim}
\usepackage[english,english]{babel}
\usepackage{amsmath,subfig}
\usepackage{amssymb,amsfonts,textcomp}
\usepackage{array}
\usepackage{supertabular}
\usepackage{hhline}
\usepackage{hyperref}
\usepackage[usenames]{color}
\hypersetup{dvips, colorlinks=true, linkcolor=black, citecolor=black, filecolor=black, urlcolor=black}
\usepackage[dvips]{graphicx}
\def\gtrsim{\lower.5ex\hbox{$\; \buildrel > \over \sim \;$}}
\usepackage{graphicx}

\newcommand{\hagn}{\mbox{{\sc \small Horizon-AGN\, }}}

\definecolor{grey}{rgb}{0.75,0.75,0.75}
\definecolor{Orange}{rgb}{1.0,0.5,0.15}
\definecolor{brown}{rgb}{0.7,0.25,0.0}
\definecolor{Pink}{rgb}{1.0,0.5,0.5}
\definecolor{darkerred}{rgb}{0.8,0,0}
\definecolor{darkerblue}{rgb}{0,0,0.8}
\definecolor{Blue}{rgb}{0,0.08,0.65}
\definecolor{Red}{rgb}{0.65,0.08,0.05}
\definecolor{Green}{rgb}{0.15,0.45,0.25}

\begin{document}

\author[Chisari et al.]{
  \parbox{\textwidth}{N. Chisari$^1$, S. Codis$^2$, C. Laigle$^{2}$, Y. Dubois$^2$, C. Pichon$^{2,3}$, J. Devriendt$^1$, A. Slyz$^1$, L. Miller$^1$, R. Gavazzi$^2$ and K. Benabed$^2$}
\vspace*{6pt}\\
\noindent
$^{1}$Department of Physics, University of Oxford, Keble Road, Oxford OX1 3RH,United Kingdom.\\
$^{2}$Institut d'Astrophysique de Paris, CNRS \& UPMC, UMR 7095, 98 bis Boulevard Arago, 75014, Paris, France.\\
$^{3}$Institute of Astronomy, University of Cambridge, Madingley Road, Cambridge, CB3 0HA, United Kingdom.
}

\date{Accepted XXXX. Received XXXX; in current form \today }

\title[Hz-AGN intrinsic alignments]{
Intrinsic alignments of galaxies in the Horizon-AGN cosmological hydrodynamical simulation}

\maketitle

\begin{abstract}
  The intrinsic alignments of galaxies are recognised as a contaminant to weak gravitational lensing measurements. In this work, we study the alignment of galaxy shapes and spins at low redshift ($z\sim 0.5$) in \hagn\!\!\!, an adaptive-mesh-refinement hydrodynamical cosmological simulation box of $100\, h^{-1}\, \rm Mpc$ a side with AGN feedback implementation. We find that spheroidal galaxies in the simulation show a tendency to be aligned radially towards over-densities in the dark matter density field and other spheroidals. This trend is in agreement with observations, but the amplitude of the signal depends strongly on how shapes are measured and how galaxies are selected in the simulation. Disc galaxies show a tendency to be oriented tangentially around spheroidals in three-dimensions. While this signal seems suppressed in projection, this does not guarantee that disc alignments can be safely ignored in future weak lensing surveys. The shape alignments of luminous galaxies in \hagn are in agreement with observations and other simulation works, but we find less alignment for lower luminosity populations. We also characterize the systematics of galaxy shapes in the simulation and show that they can be safely neglected when measuring the correlation of the density field and galaxy ellipticities.
\end{abstract}


\section{Introduction}
\label{sec:intro}

There is mounting observational evidence that galaxies are subject to `intrinsic alignments', i.e. correlations of their shapes across large separations due to tidal effects that act to align them in preferential directions with respect to one another \citep{Brown02,aubertetal04,Mandelbaum06,Hirata07,Joachimi11,Heymans13,Singh14}. These alignments have been identified as an important systematic to weak lensing measurements, with the potential to undermine its capabilities as a probe of precision cosmology if unaccounted for \citep{Hirata04,Hirata10,Bridle07,Kirk12,Krause15}. To fully extract cosmological information from weak lensing, the alignment signal needs to be mitigated or marginalized over \citep{Zhang10,Joachimi10,Joachimi10b,Troxel12}. On the other hand, there is cosmological information to be extracted from the alignment signal \citep{Chisari13,Chisari14,Schmidt15}. For these reasons, this field is emerging as an interesting avenue for improving our understanding of galaxy formation and the evolution of the large-scale structure of the Universe. For a set of comprehensive reviews on the topic of intrinsic alignments, see \citet{troxel15}, \citet{Joachimi15}, \citet{Kirk15} and \citet{Kiessling15}.

The large-scale alignment signal of the shapes of luminous red galaxies (LRGs) with the matter density field has been considered the dominant contribution to low redshift intrinsic alignments and the main source of concern for weak lensing surveys. This signal has also been measured in numerical simulations of galaxy formation \citep{Tenneti14a,Tenneti14b}. A theoretical model developed by \citet{Catelan01} suggested that elliptical galaxies, supported by random stellar motions, can be subject to large-scale tides, which stretch them along the direction of the tidal field. Alignments of LRGs are well described by this model \citep{Blazek11}.

  Disc galaxies, which have significant angular momentum, can suffer a torque from the surrounding tidal field, which correlates their orientations \citep{Catelan01,Hui02,schaefer09,Schafer15}. Hydrodynamical and $N$-body cosmological simulations have reported alignments of the spins of dark matter halos and galaxies with the cosmic web \citep{bailin&steinmetz05,Aragon07,Hahn07,sousbie08,Zhang09,Hahn10,codisetal12,Libeskind12,dubois14} and there is indeed observational evidence of galaxy spin alignments \citep{Pen00,Paz08,Jones10,andrae11,Tempel13}. Indirect evidence also comes from quasar polarizations, which have also been found to be aligned with the surrounding large-scale structure \citep{quasar14}. However, at the same time, some works have claimed no alignment of discs \citep{Slosar09,hung12}.

  Disc galaxies have higher star formation than ellipticals, hence color information is typically used as a proxy to distinguish between types and alignment mechanisms. Despite evidence of their spin alignments, the shape alignment of blue (disc) galaxies has been observed to be consistent with null \citep{WiggleZ,Heymans13}. Recently, results from an analysis of the \hagn hydrodynamical simulation\footnote{\url{http://horizon-simulation.org}} \citep{Codis14} have suggested that this correlation of the shapes of blue galaxies could be present at high significance at redshift of $z=1.2$, and it could thus become an additional contaminant to gravitational lensing. The main hypothesis behind that investigation was the use of spins as a proxy for galaxy shapes.

To elucidate the discrepancy between the non-detection in low-redshift observations of blue galaxy shape alignments and the prediction of significant spin alignments in simulations at high redshift, it is necessary to explore the relation between projected shapes and spins in the simulation in the redshift range where observational constraints are available. We therefore propose in this work to concentrate on the relationship between shapes and spins in the \hagn simulation~\citep{dubois14}. In particular, we first explore the correlations between galaxy positions and orientations to gain insights into the physical processes that give rise to intrinsic alignments. We further measure projected correlations of galaxy positions and galaxy shapes, and auto-correlations of galaxy shapes, in a way that mimics the construction of intrinsic alignment correlations from observations. Note that all results are obtained at $z=0.5$; the redshift evolution of this effect is left for future work.

In Section \ref{sec:simu}, we describe the hydrodynamical cosmological simulation used in this work to study the shape and spin alignments of galaxies, including the identification of galaxies in the simulation and the methods used to obtain their rest-frame colours. Section \ref{sec:spinshape} discusses how shapes and spins are measured from the simulation and their convergence properties. In Section \ref{sec:correl}, we present the estimators used for correlation functions between spins, positions and shapes used in this work.  We discuss both three-dimensional and two-dimensional (projected) correlations. While three-dimensional correlations give us an insight into the nature of the alignment process, the ultimate observables in photometric weak lensing surveys are the projected galaxy counts-ellipticity correlations. Section \ref{sec:results} brings together the results from the simulation and the theoretical modelling of the alignment signal. We discuss these results in relation to works in Section \ref{sec:discuss} and we conclude in Section \ref{sec:conclusion}. 

Compared to previous work by \citet{Codis14}, in this work we lift the assumption that galaxy spins are a good proxy for galaxy shapes. Instead, we present results on alignments both for spins and shapes classifying galaxies by their kinematic properties. Moreover, we concentrate in correlations between positions and orientations, rather than correlations among galaxy orientations. This allows us to neglect potential systematics arising from spurious alignment correlations due to numerical issues (see Appendix~\ref{sec:gridlock}). \citet{Tenneti14b} measured projected shape correlations of galaxies in the smoothed-particle-hydrodynamics simulation MassiveBlack-II. This work presents shape and spin correlations measured in the \hagn simulation, which has a comparable volume to MassiveBlack-II but uses a different technique: adaptive-mesh-refinement. As the properties of galaxies might be sensitive to the numerical technique adopted, the comparison of alignments measured in \hagn and MassiveBlack-II is crucial to inform predictions of intrinsic alignment contamination to future weak gravitational lensing surveys.

\section{The synthetic universe}
\label{sec:simu}

In this section, we describe the \hagn simulation (see~\citealp{dubois14} for more details) and we explain how galaxy properties relevant to this study are extracted.

\subsection{The Horizon-AGN simulation}
\label{section:numerics}

  The \hagn simulation is a cosmological hydrodynamical simulation in a box of $L=100 \, h^{-1}\rm\,Mpc$ a side. It is run adopting a standard $\Lambda$CDM cosmology compatible with the WMAP-7 cosmology~\citep{komatsuetal11}, with total matter density $\Omega_{\rm m}=0.272$, dark energy density $\Omega_\Lambda=0.728$, amplitude of the matter power spectrum $\sigma_8=0.81$, baryon density $\Omega_{\rm b}=0.045$, Hubble constant $H_0=70.4 \, \rm km\,s^{-1}\,Mpc^{-1}$, and $n_s=0.967$. There are $1024^3$ dark matter (DM) particles in the box, with a resulting DM mass resolution of $M_{\rm DM, res}=8\times 10^7 \, \rm M_\odot$.
  
The adaptive-mesh-refinement (AMR) code {\sc ramses}~\citep{teyssier02} has been used to run the simulation with $7$ levels of refinement up to $\Delta x=1\, \rm kpc$. The refinement scheme follows a quasi-Lagrangian criterion. A new refinement level is triggered if the number of DM particles in a cell is more than $8$ or if the total baryonic mass in a cell is $8$ times the initial DM mass resolution.

Star formation is modelled following a Schmidt law: $\dot \rho_*= \epsilon_* {\rho / t_{\rm ff}}\, ,$ where $\dot \rho_*$ is the star formation rate density, $\rho$ is the gas density, $\epsilon_*=0.02$~\citep{kennicutt98, krumholz&tan07} the constant star formation efficiency, and $t_{\rm ff}$ the local free-fall time of the gas. We allow star formation wherever the Hydrogen gas number density exceeds $n_0=0.1\, \rm H\, cm^{-3}$ according to a Poisson random process~\citep{rasera&teyssier06, dubois&teyssier08winds} with a stellar mass resolution of $M_{*, \rm res}=\rho_0 \Delta x^3\simeq 2\times 10^6 \, \rm M_\odot$.

Gas cooling occurs by means of H and He cooling down to $10^4\, \rm K$ with a contribution from metals \cite{sutherland&dopita93}. Following~\cite{haardt&madau96}, heating from a uniform UV background is implemented after the reionization redshift $z_{\rm reion} = 10$. Metallicity is modeled as a passive variable of the gas, which varies according to the injection of gas ejecta from supernovae explosions and stellar winds. We model stellar feedback using a \citet{salpeter55} initial mass function with a low-mass (high-mass) cut-off of $0.1\, \rm M_{\odot}$ ($100 \, \rm M_{\odot}$). 
In particular, the mechanical energy from supernovae type II and stellar winds follows the prescription of {\sc starburst99}~\citep{leithereretal99, leithereretal10}, and the frequency of type Ia supernovae explosions is taken from~\cite{greggio&renzini83}.

Active Galactic Nuclei (AGN) feedback is modelled according to~\cite{duboisetal12agnmodel}. We use a Bondi-Hoyle-Lyttleton accretion rate onto black holes, given by $\dot M_{\rm BH}=4\pi \alpha G^2 M_{\rm BH}^2 \bar \rho / (\bar c_s^2+\bar u^2) ^{3/2},$ where $M_{\rm BH}$ is the black hole mass, $\bar \rho$ is the average gas density, $\bar c_s$ is the average sound speed, $\bar u$ is the average gas velocity relative to the black hole velocity, and $\alpha$ is a dimensionless boost factor. This is given by $\alpha=(\rho/\rho_0)^2$ when $\rho>\rho_0$ and $\alpha=1$ otherwise~\citep{booth&schaye09} in order to account for our inability to capture the colder and higher density regions of the interstellar medium. The effective accretion rate onto black holes is not allowed to exceed the Eddington accretion rate: $\dot M_{\rm Edd}=4\pi G M_{\rm BH}m_{\rm p} / (\epsilon_{\rm r} \sigma_{\rm T} c),$ where $\sigma_{\rm T}$ is the Thompson cross-section, $c$ is the speed of light, $m_{\rm p}$ is the proton mass, and $\epsilon_{\rm r}$ is the radiative efficiency, assumed to be equal to $\epsilon_{\rm r}=0.1$ for the \cite{shakura&sunyaev73} accretion onto a Schwarzschild black hole. Two different modes of AGN feedback are implemented: the \emph{radio} mode, operating when $\chi=\dot M_{\rm BH}/\dot M_{\rm Edd}< 0.01$, and the \emph{quasar} mode, active otherwise (see~\citealp{dubois14} for details).

\subsection{Galaxy catalogue}
\label{section:postprocess}

Galaxies are identified in each redshift snapshot of \hagn using the AdaptaHOP finder~\citep{aubertetal04}, which relies directly on the distribution of stellar particles. Twenty neighbours are used to compute the local density of each particle. To select relevant over-densities, we adopt a local threshold of $\rho_{\rm t}=178$ times the average total matter density. Note that the galaxy population does not depend sensitively on the exact value chosen for this threshold. Our specific choice reflects the fact that the average density of galaxies located at the centre of galaxy clusters is comparable to that of the dark matter. The force softening is of approximately $10$ kpc. Below this minimum size, substructures are treated as irrelevant.

Only galactic structures identified with more than 50 star particles are included in the mock catalogues. This enables a clear identification of galaxies, including those in the process of merging. At $z=0.5$, there are $\sim 146 \, 000$ objects in the galaxy catalogue, with masses between $1.7\times10^{8}$ and $2.3\times10^{12}\, \rm M_{\odot}$. The galaxy stellar masses quoted in this paper should be understood as the sum over all star particles that belong to a galaxy structure identified by AdaptaHOP. 

We compute the absolute AB magnitudes and rest-frame colours of the mock galaxies using single stellar population models from~\cite{bruzual&charlot03} adopting a Salpeter initial mass function. Each stellar particle contributes a flux per frequency that depends on its mass, metallicity and age. The sum of the contribution of all star particles is passed through $u$, $g$, $r$, and $i$ filter bands from the {\it Sloan Digital Sky Survey} \citep[SDSS,][]{Gunn06}. Fluxes are expressed as rest-frame quantities (i.e. that do not take into account the redshifting of spectra) and, for simplicity, dust extinction is neglected. Once all the star particles have been assigned a flux in each of the colour channels, we build the 2D projected maps for individual galaxies (satellites are excised with the galaxy finder). The sum of the contribution of their stars yields the galaxy luminosity in a given filter band.

\section{Spins and shapes of galaxies}
\label{sec:spinshape}

To assign a spin to the galaxies, we compute the total angular momentum of the star particles which make up a given galactic structure relative to the centre of mass. We can therefore write the intrinsic angular momentum vector $\mathbf{L}$ or spin of a galaxy as
\begin{equation}
\label{eq:spindef}
  \mathbf{L} = \sum_{ n=1}^{N} m^{( n)} \mathbf{x}^{(n)} \times \mathbf{v}^{(n)} \, ,
\end{equation}
where $n$ denotes each stellar particle of mass $m^{ (n)}$, position $\mathbf{x}^{ (n)}$ and velocity $\mathbf{v}^{(n)}$ relative to the center of mass of that galaxy. The total stellar mass of a galaxy is given by~$M_*= \sum_{n=1}^{N} m^{ (n)}$. The specific angular momentum is ${\bf j}={\bf L}/M_*$.

For each galaxy, we also obtain its $V/\sigma$, stellar rotation versus dispersion, that we measure from their 3D distribution of velocities. $V/\sigma$ is a proxy for galaxy morphology: low values of this physical quantity indicate that a galaxy is more elliptical, pressure-supported by random stellar motions; high values of $V/\sigma$ suggest a disc-like galaxy, with stellar rotation predominantly on a plane.
We first compute the total angular momentum (spin) of stars in order to define a set of cylindrical spatial coordinates ($r$, $\theta$, $z$), with the $z$-axis oriented along the galaxy spin. The velocity of each individual star particle is decomposed into cylindrical components $v_{r}$, $v_{\theta}$, $v_z$, and the rotational velocity of a galaxy is $V=\bar v_{\theta}$, the mean of $v_{\theta}$ of individual stars. The velocity dispersion of each velocity component $\sigma_{r}$, $\sigma_{\theta}$, $\sigma_z$ is computed and used for the average velocity dispersion of the galaxy $\sigma^2=(\sigma_{r}^2+\sigma_{\theta}^2+\sigma_z^2)/3$.

\begin{figure}
\centering
\includegraphics[width=0.45\textwidth]{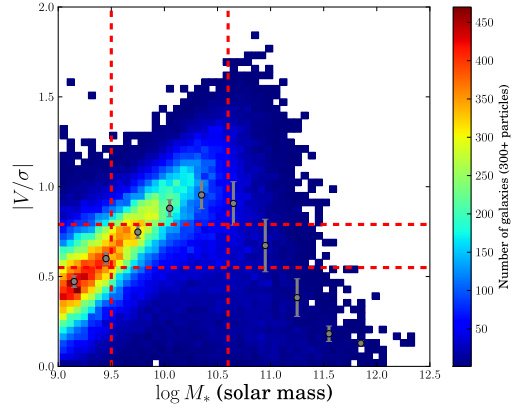}
\caption{$V/\sigma$ as a function of mass for the galaxies used in this work. The gray circles represent the median and the variance in $10$ logarithmic bins of stellar mass. Dwarf and massive galaxies are pressure-supported, while intermediate-mass have a mixture of rotation and pressure support. The vertical red dashed lines represent the cuts corresponding to the different mass bins as in \citet{Codis14}. The horizontal red dashed lines represent our cuts in $V/\sigma$, chosen such that there is approximately the same number of galaxies in each $V/\sigma$ bin.}
\label{fig:vsig}
\end{figure}

Figure~\ref{fig:vsig} shows the distribution of $V/\sigma$ for galaxies in the \hagn simulation. We see an increase of the ratio $V/\sigma$ from low-mass $M_*\simeq 10^9\, \rm M_\odot$ to intermediate-mass $M_*\simeq 2\times 10^{10}\, \rm M_\odot$  galaxies, for which this median of the ratio peaks at $1$. For massive galaxies, above $2\times 10^{10}\, \rm M_\odot$, the $V/\sigma$ ratio decreases with stellar mass. Dwarf and massive galaxies have low $V/\sigma$ and hence are pressure-supported, while intermediate-mass galaxies have wide-spread $V/\sigma$ values and are thus a hybrid population of rotation- and pressure-supported galaxies.

We also measure the inertia tensor of a galaxy to characterize its three dimensional shape. This tensor is given by 
\begin{equation}\label{eq:inerdef}
I_{ ij}=\frac{1}{M_*}\sum_{n=1}^{N} m^{(n)} x^{(n)}_{i} x^{(n)}_{j}\, .
\end{equation}
and it is then diagonalised to obtain the eigenvalues $\lambda_1\le \lambda_2 \le \lambda_3$ and the corresponding unit eigenvectors $\mathbf{u}_1$, $\mathbf{u}_2$ and $\mathbf{u}_3$ (respectively minor, intermediate and major axis of the ellipsoid). Analogously, we will explore the differences in measuring galaxy shapes using Equation (\ref{eq:inerdef}) compared to the {\it reduced} inertia tensor, which is defined by
\begin{equation}\label{eq:reduceddef}
{\tilde I}_{ ij}=\frac{1}{M_*}\sum_{n=1}^{N} m^{(n)} \frac{x^{(n)}_{i} x^{(n)}_{j}}{r_{n}^2},
\end{equation}
where $r_{n}^2$ is the three dimensional distance for the stellar particle $n$ to the center of mass of the galaxy. 
The reduced inertia tensor is a closer representation of the shape of a galaxy as measured for weak gravitational lensing measurements, where the inner (and more luminous) region is upweighted with respect to the outskirts. For the reduced inertia tensor, \citet{Tenneti14a} found that the iterative procedure reduces the impact of the spherically symmetric $r^{-2}$ weights, yielding shapes that are not as round as for the non-iterative procedure. We will consider here the simple and the reduced inertia tensor cases, and we expect that the results from applying iterative procedures to define the galaxy shapes \citep{Schneider12,Tenneti14a} will lie between the two cases considered.

Projected shapes are obtained by summing over $i,j=1,2$ in Equations (\ref{eq:inerdef}) and (\ref{eq:reduceddef}), and the semiminor and semimajor axes are correspondingly defined by the eigenvectors of the projected inertia tensors. The axis ratio of the galaxy, $q=b/a$, is defined from the eigenvalues of the projected inertia tensor as the ratio of the minor to major axes ($b=\sqrt{\lambda_b},a=\sqrt{\lambda_a}$, where $\lambda_a$ is the largest eigenvalue and $\lambda_b$, the smallest). The components of the complex ellipticity, typically used in weak lensing measurements, are given by
\begin{equation}
  (e_+,e_\times) = \frac{1-q^2}{1+q^2}[\cos(2\phi),\sin(2\phi)]\,,
  \label{eq:complexe}
\end{equation}
where $\phi$ is the orientation angle, $+$ indicates the radial component of the ellipticity and $\times$ is the $45\deg$-rotated component. Intrinsic alignments typically manifest themselves as a net average $e_+$ ellipticity around over-densities. The total ellipticity of a galaxy is thus $e=\sqrt{e_+^2+e_\times^2}$. In this work, radial alignments have negative $e_+$ and tangential alignments (as expected from weak gravitational lensing) correspond to positive $e_+$. No net correlation of the $\times$ component with the matter density field is expected, and this statistic is thus used to test for systematics in the ellipticity measurement procedures. We discuss other systematics tests of ellipticity correlations in detail in Appendix~\ref{sec:gridlock}. 

\subsection{Convergence tests}
\label{subsec:convergence}

Poorly resolved galaxies in the simulation are subject to uncertainties in their shapes and orientations. We identify two sources of uncertainties. First, there is the inherent variance in the shapes that arises from the choice of particles used in the shape computation. We call this measurement noise, $\sigma_{\rm meas}$, in analogy to the shape measurement from galaxy images for weak lensing. Second, there is a bias in the ellipticity and orientation measurement associated to the resolution of a galaxy, $\sigma_{\rm res}$. To define the minimum number of particles needed to obtain the shape of a galaxy, we compare these two uncertainties to the shape noise that arises from the dispersion in the intrinsic distribution of the shapes, $\sigma_e$.

\begin{figure}
\centering
\includegraphics[width=0.5\textwidth]{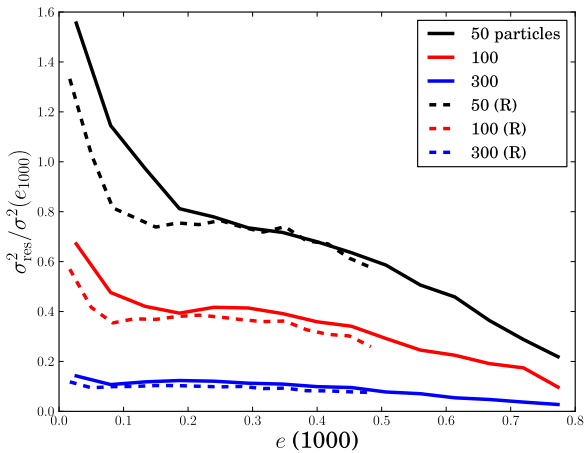}
\caption{Resolution bias, $\sigma_{\rm res}^2 = \langle |e_{50,100,300}-e_{1000}|^2 \rangle/\sigma^2(e_{1000})$, as a function of the ellipticity obtained from the $1000$ particle subsample for galaxies with more than 1000 particles. The colours indicate the number of stellar particles in the subsample: $>50$ (black), $>100$ (red) and $>300$ (blue). The solid lines represent the results for the simple inertia tensor and the dashed lines, for the reduced inertia tensor. Subsampling has a larger impact on the simple inertia tensor due to the fact that particles in the outskirts contribute with equal weights as particles in the center.}
\label{fig:sigmares}
\end{figure}

\begin{figure}
\centering
\includegraphics[width=0.5\textwidth]{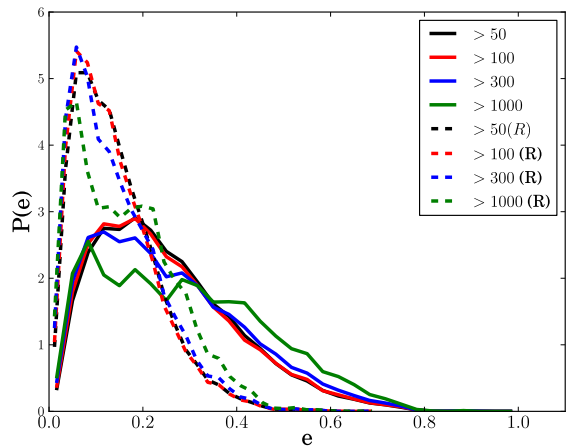}
\caption{Distribution of ellipticities for galaxies with $>50$ (black), $>100$ (red), $>300$ (blue) and $>1000$ particles (green) for the simple (solid) and the reduced (dashed) inertia tensors. The shapes obtained with the reduced inertia tensor are rounder than those obtained with the simple inertia tensor.}
\label{fig:edist}
\end{figure}

The resolution bias is determined by randomly subsampling stellar particles in each galaxy with $>1000$ particles. We compare the ellipticity measured from random subsamples of $50,100,300$ and $1000$ stellar particles for those galaxies. The results are shown in Figure~\ref{fig:sigmares} and they suggest that a minimum number of $300$ particles in each galaxy has to be required in order to guarantee that the bias in the ellipticity is an order of magnitude below the shape noise: $\sigma_{\rm res}^2\lesssim 0.1\,\sigma_e^2$. The distributions of galaxy ellipticities are shown in Figure~\ref{fig:edist}. 

\begin{figure}
\centering
\includegraphics[width=0.5\textwidth]{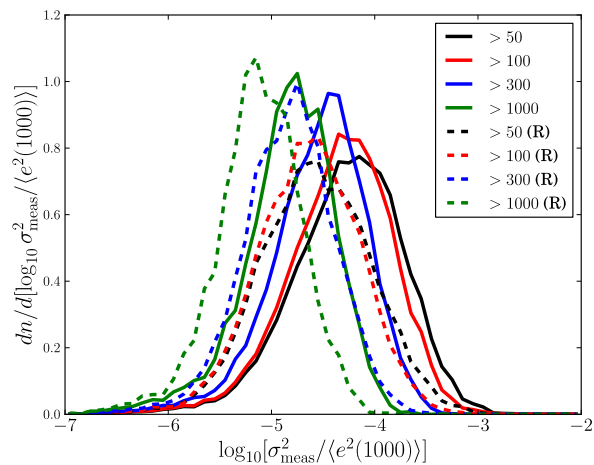}
\caption{Distribution of measurement uncertainties in the ellipticities for galaxies with $>50$ (black), $>100$ (red), $>300$ (blue) and $>1000$ particles (green) for the simple (solid) and the reduced (dashed) inertia tensors. The shapes obtained with the reduced inertia tensor have significantly smaller uncertainties. These uncertainties are insignificant compared to the resolution bias and to the shape noise in the sample. }
\label{fig:sigmameas}
\end{figure}

We determine $\sigma_{\rm meas}$ by bootstrap resampling ($100$ times) the stellar particles used for defining the inertia tensor from the overall population of stellar particles that make up each galaxy. We compute this uncertainty for galaxies with $>50$, $>100$, $>300$ and $>1000$ particles. Figure~\ref{fig:sigmameas} shows the distribution of uncertainty in the shape measurement (compared to the rms ellipticity of the galaxy sample with $1000$ particles) for each minimum number of particles considered. The solid lines correspond to the shape measured using the simple inertia tensor and the dashed lines, to those measured using the reduced inertia tensor. The uncertainties are typically below the $10^{-3}$ level and they are smaller in the case of the reduced inertia tensor. Resampling has a larger impact on the simple inertia tensor, since there are few particles in the outskirts of each galaxy, and these contribute equally to the shape measurement as those particles in the central part. Nevertheless, $\sigma_{\rm meas}$ is very small compared to the shape noise and to the resolution bias, and hence can be neglected. Notice that in our modelling we do not include surface brightness cuts, attenuation by dust, the effect of the atmosphere or convolution by telescope optics. Also, the resampling of stellar particles is carried out over all stellar particles, which constitute a correlated data set. All of these could have a significant impact in the estimation of $\sigma_{\rm meas}$ but their modelling is outside the scope of this work.

\begin{figure}
\centering
\includegraphics[width=0.5\textwidth]{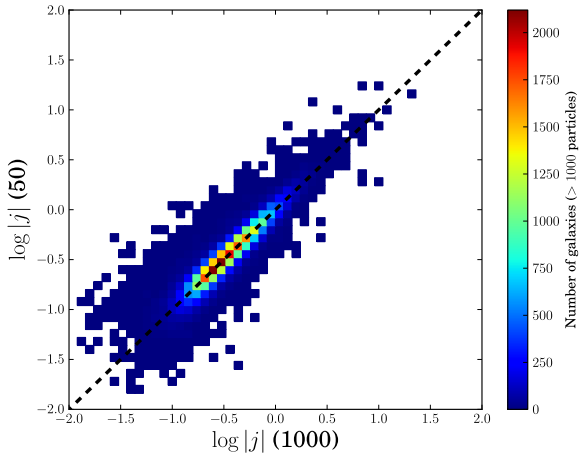}
\caption{Correlation between the logarithm of the specific angular momentum, $j=|{\bf L}|/M_*$, when measured with $50$ ($y$-axis) and $1000$ ($x$-axis) particles, for galaxies in the simulation with $>1000$ particles.}
\label{fig:j50_j1000}
\end{figure}

We also estimated the impact of subsampling and resampling on the measurement of the specific angular momentum of the galaxies. We found that the measurement noise is insignificant and that the specific angular momentum is less sensitive to the number of particles in the subsample than the projected shapes. While subsampling introduces significant scatter in the measurement of the specific angular momentum, there is a strong correlation between this quantity measured with $50$, $100$, $300$ and $1000$ stellar particles. This correlation is shown for $50$ versus $1000$ particles in Figure~\ref{fig:j50_j1000}.

\begin{figure}
\centering
\includegraphics[width=0.5\textwidth]{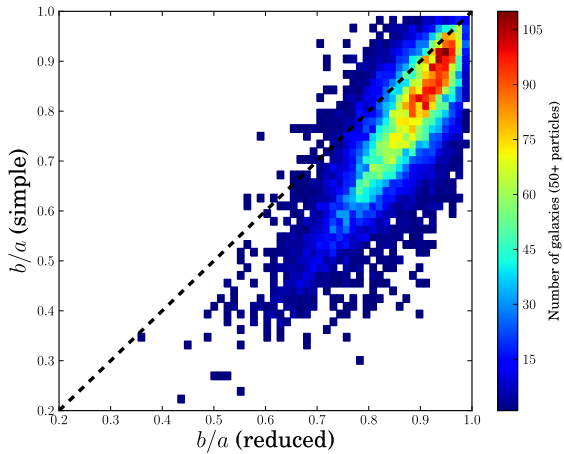}
\caption{Axis ratios for galaxies with $>50$ particles obtained with the simple inertia tensor ($y$-axis) an the reduced inertia tensor ($x$-axis). Colours indicate the number of galaxies in each bin. While there is a clear correlation between the estimates obtained with the two different methods, the reduced inertia tensor tends to produced rounder shapes.}
\label{fig:simplereduced}
\end{figure}

We conclude that the minimum number of particles required to perform spin and shape measurements is determined by the requirements placed on the projected shapes. A minimum of $300$ particles is needed for the uncertainty in projected shape to be at least one order of magnitude smaller than the shape noise. From this section onwards, we will work only with galaxies that have more than $300$ stellar particles in the simulation. Notice that \citet{Tenneti14b} adopted a cut on $1000$ stellar particles, while \citet{Velliscig15} adopt similar cuts as in our work on the number of stellar particles. Figure~\ref{fig:simplereduced} shows a comparison between the axis ratios of galaxies in the simulation using the simple ($y$-axis) and the reduced ($x$-axis) inertia tensors for galaxies with more than $300$ stellar particles. The use of the reduced inertia tensor clearly results in rounder shapes as a result of up-weighting the inner regions of galaxies. 


\section{Correlation functions}
\label{sec:correl}

\subsection{Correlations in three dimensions}
\label{sec:corr3d}

Three dimensional correlations of galaxy shapes and spins can give interesting insights into the formation processes leading to alignments. In this section, we define the correlation functions of galaxy {\it orientations} in three dimensional space. We define the orientation-separation correlation as a function of comoving separation $r$,
\begin{equation}
\eta_e(r) = \langle |\hat{\bf r}\cdot \hat{\bf e}({\bf x}+{\bf r})|^2\rangle - {1}/{3}\,,
\label{eq:etaER}
\end{equation}
where $\hat{\bf e}$ is the unit eigenvector of the inertia tensor pointing in the direction of the minor axis and $\hat{\bf r}$ is the unit separation vector. A positive correlation indicates a tendency for the separation vector and the minor axis of a galaxy to be parallel, hence for the galaxy to be elongated tangentially with respect to another galaxy. Notice that gravitational lensing produces a similar correlation for galaxy pairs separated by large distances along the line of sight. A negative correlation corresponds to a preferential perpendicular orientation of the minor axis with respect to the separation vector, resulting in a net radial orientation of galaxy shapes around other galaxies. Our treatment of galaxy orientations corresponds to representing the galaxy with the normal of its orientation plane and does not take into account the intrinsic dispersion in ellipticities. In other words, both spheroidals and discs are infinitely thin in this approach, and the plane that represents them points in the direction perpendicular to $\hat{\bf e}$. Alternatively, we will also measure the analogous correlation of separation vectors with spins, $\hat{\bf s}$, defined from the angular momentum of a galaxy, $\eta_s(r)$.

Grid-locking (studied in detail in Appendix~\ref{sec:gridlock}) manifests itself as a correlation between spin and shapes with the directions of the simulation box. As a result, spin and ellipticity auto-correlations (presented in Appendix~\ref{sec:auto}) can be contaminated by this effect, as two nearby galaxies formed from the same stream of infalling gas will be correlated among themselves and with the grid in which forces are evaluated and the effect can be enhanced at small distances because of the correlation length of the force field. On the contrary, we demonstrate that position-shape, position-spin and position-orientation correlations presented in the main body of this work are not affected. 

\subsection{Projected correlations}
\label{sec:projected}

Three dimensional information of galaxy shapes and orientations is costly and near-term imaging surveys will be limited to correlations between projected shapes in broad tomographic bins. In this section, we describe projected statistics of alignments. First, we determine the orientation and ellipticity (Equation~ \ref{eq:complexe}) of a galaxy from the projected inertia tensor. We obtain the real-space correlation function of galaxy positions and shapes projecting along one of the coordinates of the simulation box in a way that mimics the construction of Landy-Szalay estimators \citep{LS93} in imaging surveys.

We label the dataset of tracers of the density field by $D$, the set of galaxies with ellipticities by $S_{+}$ (for the tangential ellipticity component, and analogously for the $\times$ component) and the set of random points by $R$. We define the correlation function of galaxy positions and ellipticities, $\xi_{g,+}(r_p,\Pi)$ as a function of projected separation in the sky, $r_p$, and along the line of sight, $\Pi$. This function is estimated from the sample of galaxies by using a modified Landy-Szalay estimator~\citep{WiggleZ}
\begin{align}
\xi_{g,+}(r_p,\Pi) &= \frac{S_+D - S_+R}{RR}\,,\label{eq:xigp}\\
S_+D &= \sum_{(r_p,\Pi)} \frac{e_{+,j}}{2\mathcal{R}}\,\label{eq:splusD},
\end{align}
where $\mathcal{R}$ is the responsivity factor \citep{Bernstein02}, $\mathcal{R}=1-\langle e^2 \rangle$, $e_{+,j}$ is the tangential/radial component of the ellipticity vector of galaxy $j$ and the sum is over galaxy pairs in given bins of projected radius and line of sight distance. $S_+R$ is the sum of galaxy ellipticities around random points. In the previous expression, we are implicitly assuming that there is a single random sample, a procedure which is valid when the galaxy sample and the shape sample coincide. For the random sample, we use a set of points uniformly distributed in the simulation box with $10$ times the density of the galaxy sample. For cross-correlations of different samples, generalization of Equation~(\ref{eq:xigp}) is straight-forward \citep[e.g.][]{Singh14}.

This correlation function is then projected along the line of sight by integration between $-\Pi_{\rm max}<\Pi<\Pi_{\rm max}$\footnote{Notice that for simplicity we do not include the effect of peculiar velocities in transforming galaxy redshifts into distance along the line of sight, nor do we incorporate photometric redshift uncertainties. Both effects decrease the significance of intrinsic alignment correlations; however, in this work, we are interested in extracting as much information as possible from the simulation and we opt to avoid diluting the alignment signal by including these effects in the modelling.},
\begin{equation}
  w_{g+}(r_p) = \int_{-\Pi_{\rm max}}^{\Pi_{\rm max}} \textrm{d}\Pi\,\xi_{g+}(r_p,\Pi)\,,
  \label{eq:wgplusdef}
\end{equation}
where we take $\Pi_{\rm max}$ to be half the length of the simulation box. We are also interested in the correlation between the density field and the galaxy shapes, $w_{\delta+}(r_p)$, for which the galaxy positions are replaced with the positions of randomly sampled DM particles in the simulation box. 

Two sources of uncertainties can be identified when measuring projected correlation functions of positions and shapes. The first one is shot noise coming from the tracers of the density and shear field. The second source of uncertainty is cosmic variance from the limited volume of the simulation box. Given that the box is $100 \, h^{-1}\, \rm Mpc$ 
on each side, we limit the measurement of correlations to up to a quarter the length of the box size, $L$. However, this does not guarantee that we will be free from cosmic variance for the measured separations. The uncertainties in the projected correlation functions, including both shape noise and cosmic variance from modes within the box, can be obtained from jackknife resampling over the simulation box \citep{Hirata04b,Mandelbaum06b}. We divide the simulation box in cubes of $L/3$ a side. In each jackknife iteration we remove the tracers corresponding to one of the cubes and compute a new estimate of the correlation. This allows us to estimate the variance in the projected correlations including the effect of cosmic variance. By contrast, the shape noise-only variance underestimates the jackknife variance by a factor of $1-4$ on small scales ($<\,1 \, h^{-1}\, \rm Mpc$) and up to approximately an order of magnitude on the largest scales probed in this work. In comparison, the jackknife accounts for covariance between $\Pi$ bins and for the finite volume of the simulation box. Nevertheless, due to the limited size of the box, it is expected that the error bars could still be underestimated on large scales. We have also confirmed that using more jackknife regions with $L/4$ length a side does not alter the main conclusions of this manuscript.

\section{Results}
\label{sec:results}

\subsection{Relative orientations in three dimensions}
\label{sec:align3d}

In this section, we characterize three dimensional correlations of galaxy positions and orientations at $z\simeq 0.5$ to shed light into the physical processes that can lead to alignments. Since three dimensional shapes are costly (see \citealt{Huff13} for a potential method), we study on-the-sky galaxy-ellipticity correlations in Section \ref{sec:resproj}.

\begin{figure*} 
  \centering
  \includegraphics[width=0.33\textwidth]{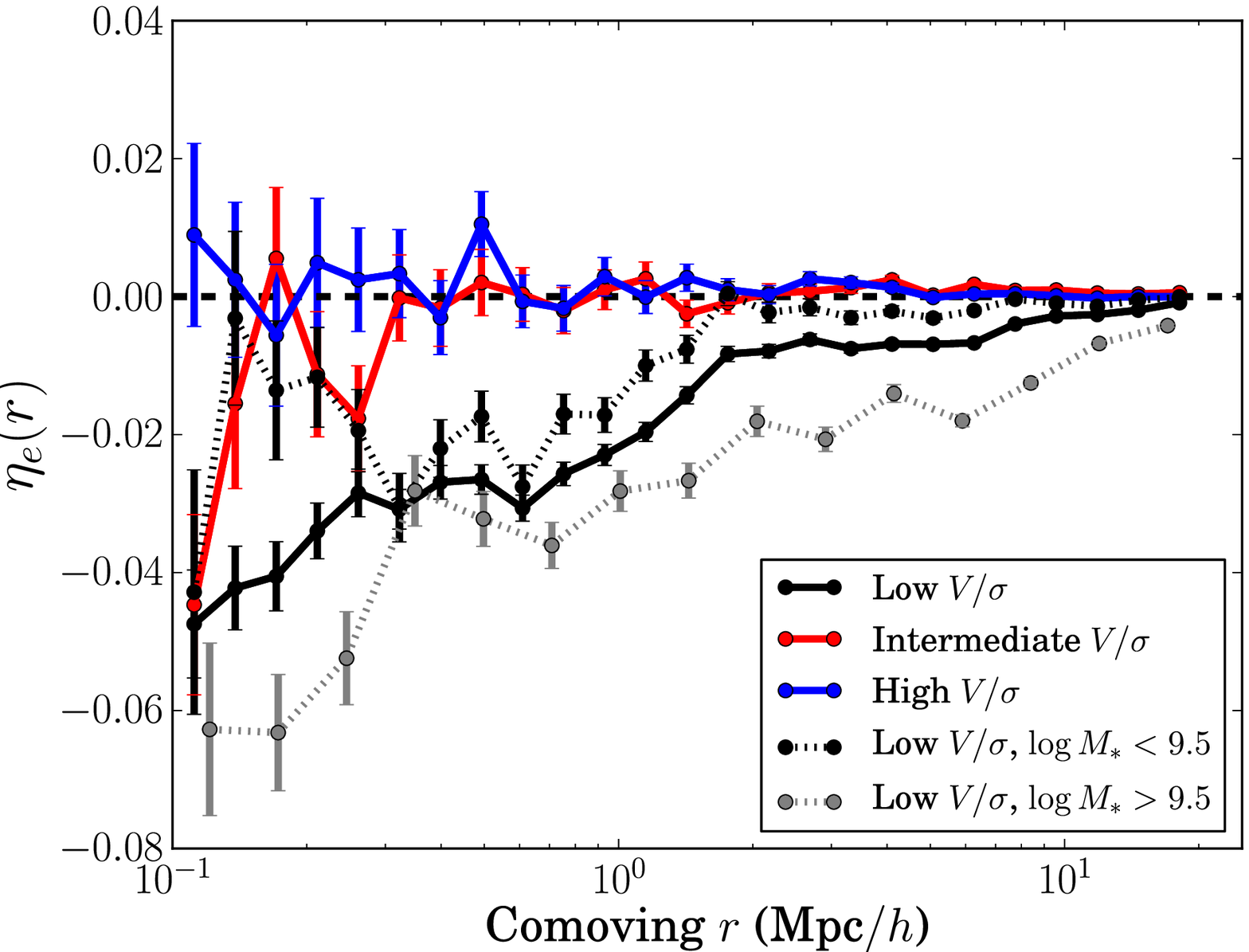}
  \includegraphics[width=0.33\textwidth]{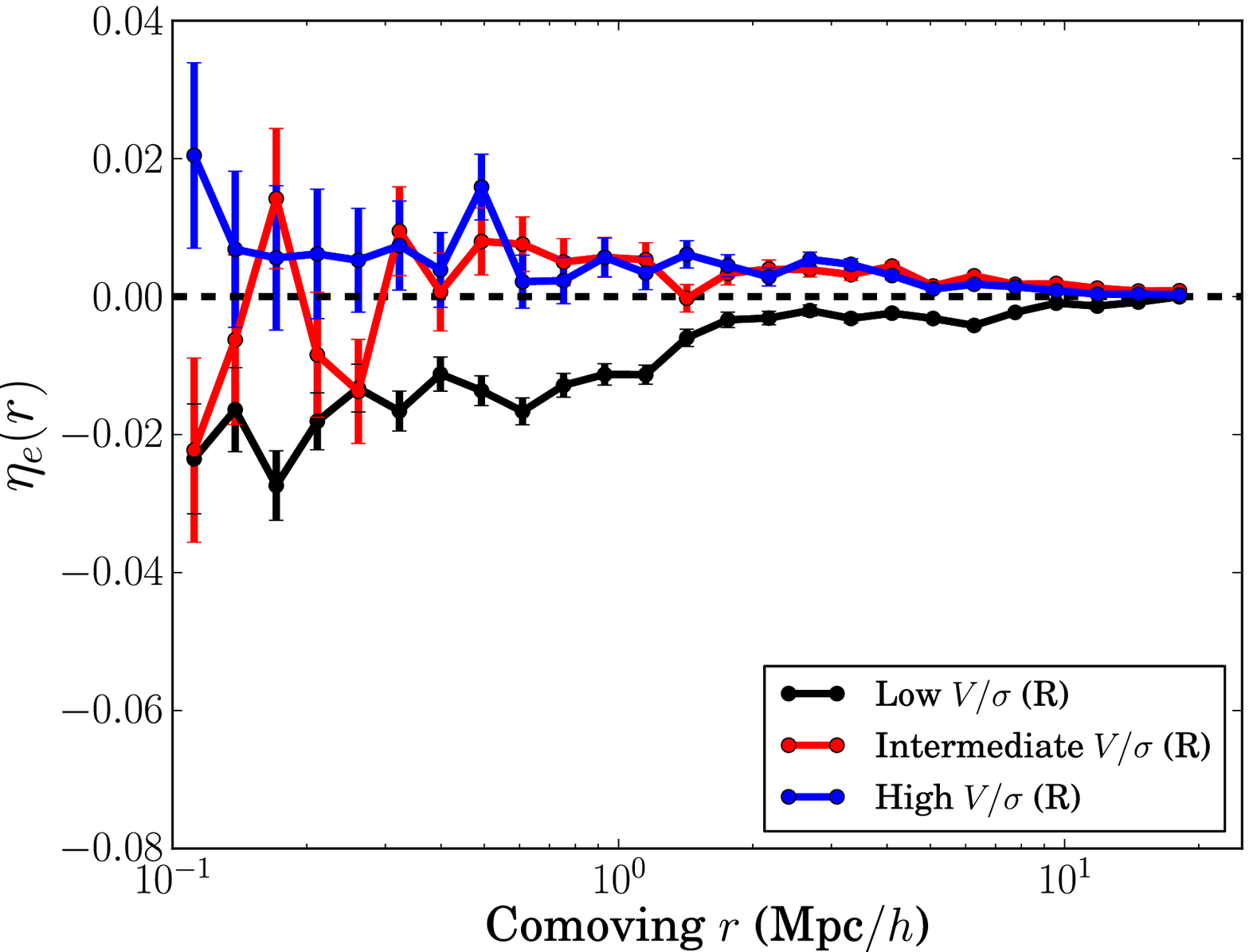}
  \includegraphics[width=0.33\textwidth]{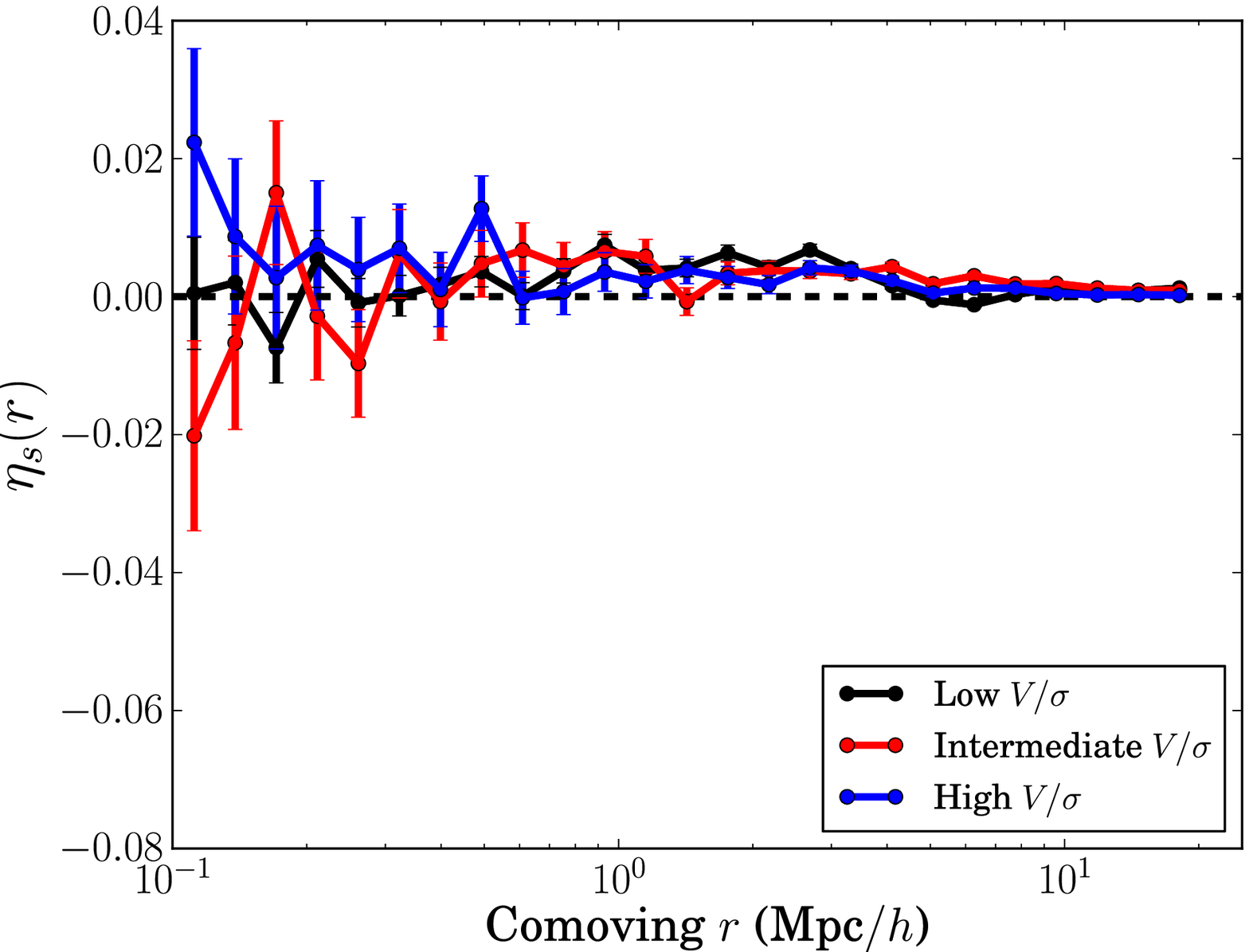}
\caption{Correlations between the minor axis ($\eta_e$, left for simple inertia tensor and middle panel for reduced inertia tensor) or spin ($\eta_s$, right panel) and separation vector as a function of comoving separation. The sample of galaxies is divided into three bins by their dynamical classification: $V/\sigma<0.55$, $0.55<V/\sigma<0.79$ and $V/\sigma>0.79$. Galaxies with low $V/\sigma$ show a significant shape-separation correlation. In the left panel, the dotted lines show the alignment signal for those galaxies split by the mean mass of that population: $\log M_*<9.5$ (black) and $\log M_* > 9.5$ (gray). Both low and high mass populations have a contribution, and higher mass galaxies have stronger alignments.}
\label{fig:etavsig}
\end{figure*}
\begin{figure*} 
\centering
\includegraphics[width=0.33\textwidth]{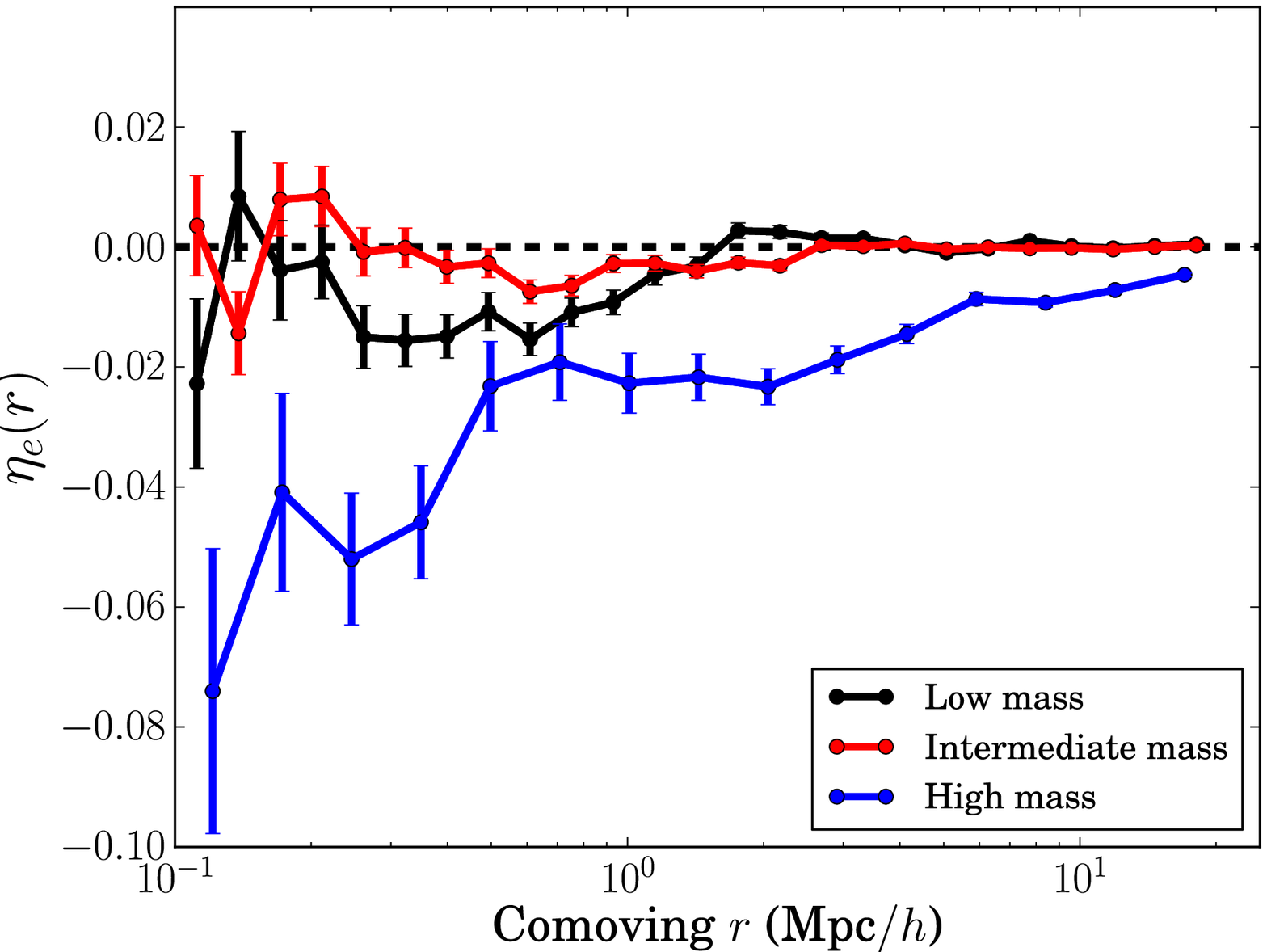}
\includegraphics[width=0.33\textwidth]{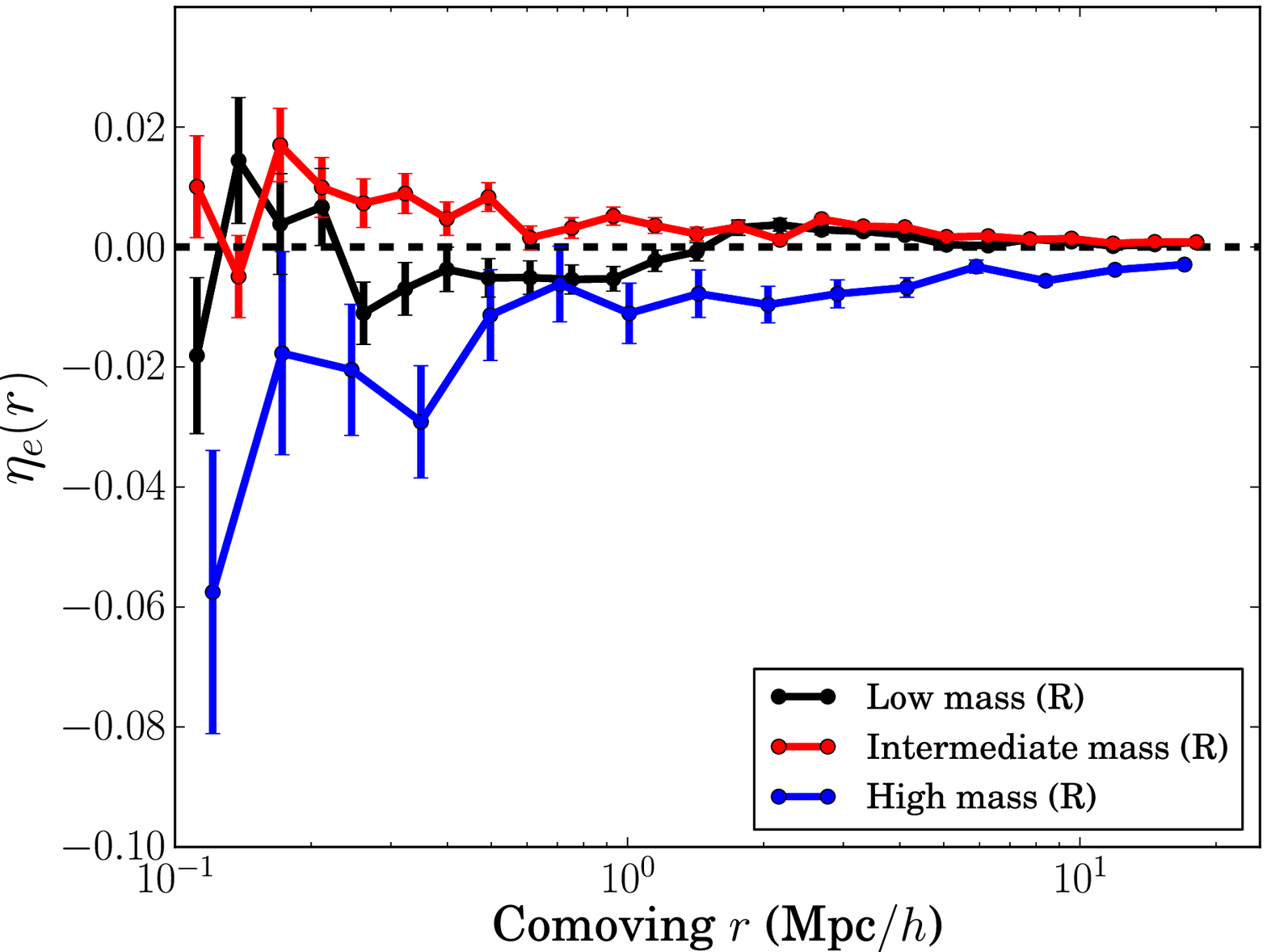}
\includegraphics[width=0.33\textwidth]{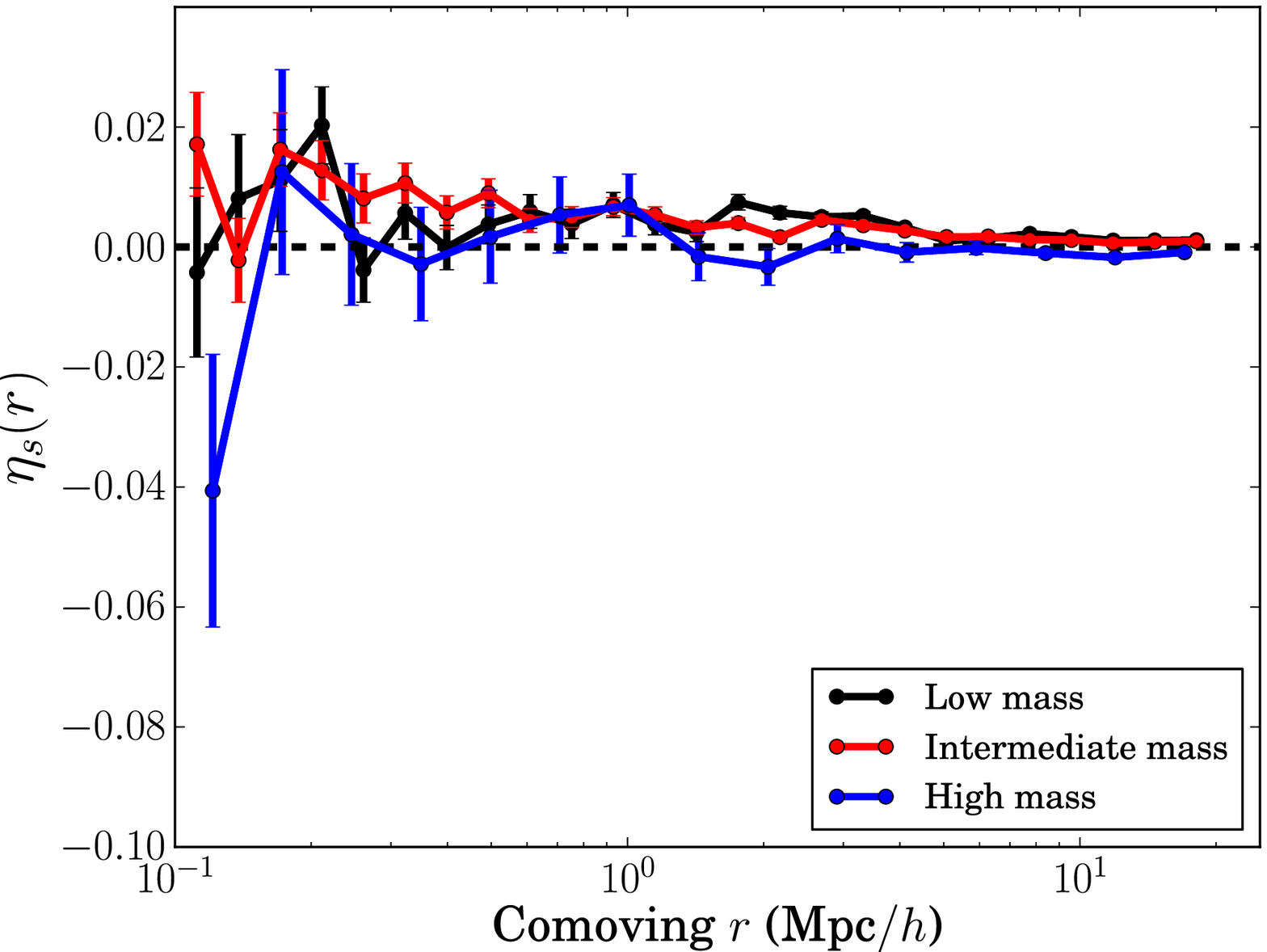}
\caption{Correlation between the direction of the the minor axis ($\eta_e$) obtained from the simple inertia tensor (left panel) and from the reduced inertia tensor (middle panel) or spin of a galaxy ($\eta_s$, right panel) with the separation vector to a galaxy within the same mass-selected sample as a function of comoving separation. High mass galaxies show a clear radial alignment trend towards other high mass galaxies.}
\label{fig:dominor}
\end{figure*}

We use all galaxies in the simulation with $>300$ stellar particles divided into three subpopulations by their $V/\sigma$, which is obtained as described in Section \ref{sec:spinshape}. This allows us to determine whether the alignment trends depend on the dynamical properties of the galaxies. Disc galaxies are represented by large values of $V/\sigma$, while spheroidals have low $V/\sigma$. As mentioned in Section~\ref{sec:intro}, different alignment mechanisms are expected to act on these different populations \citep{Catelan01}. Disc galaxies are expected to become aligned with the large-scale tidal field due to torques in their angular momentum, while ellipticals might suffer stretching or accretion along preferential directions determined by large-scale tides.

The three $V/\sigma$ bins are constructed to host $1/3$ of the galaxy population in each bin ($V/\sigma<0.55$, $0.55<V/\sigma<0.79$ and $V/\sigma>0.79$). The left panel of Figure~\ref{fig:etavsig} shows the three-dimensional correlation function of minor axis, from the simple inertia tensor, and separation vector. Error bars correspond to the standard error on the mean in each bin. We are only interested in a qualitative comparison in this section and we thus neglect cosmic variance except when explicitly mentioned. We observe that the minor axis orientation-position correlation is only significant in the low $V/\sigma$ bin, suggesting that only galaxies with spheroidal dynamics are subject to shape alignments. The results imply that these galaxies are elongated pointing towards each other, in agreement with the qualitative behaviour expected from the tidal alignment model \citep{Catelan01}. (Notice that, by construction, the fraction of spheroidal galaxies at this redshift is $1/3$.) Galaxies with $V/\sigma>0.55$ do not have correlated positions and shapes. Moreover, low $V/\sigma$ have a lower degree of correlation between the direction of their minor axis and the direction of their spin, and this is a monotonic function of $V/\sigma$. Finally, we have split the low $V/\sigma$ population of galaxies by the mean stellar mass; we find that both low ($\log M_* <9.5$, $\sim20,000$ galaxies) and high mass ($\log M_* >9.5$, $\sim8,000$ galaxies) galaxies are subject to alignments, and that high mass galaxies are more strongly aligned. Spin alignments are not very significant among these populations, although there seems to be a small trend for galaxies being oriented perpendicularly to the separation, i.e., tangentially around other galaxies. This also seen in the reduced inertia tensor of low and intermediate $V/\sigma$ galaxies.

We also divide the galaxy population into three bins of mass: low mass ($10^{9}<M_*<10^{9.5}\, \rm M_\odot$, $\sim 30,000$ galaxies), intermediate mass ($10^{9.5}<M_*<10^{10.6}\, \rm M_\odot$, $\sim 40,000$ galaxies) and high mass ($M_*>10^{10.6} \, \rm M_\odot$, $\sim 7,000$ galaxies). The low mass limit is determined by the required threshold in the number of stellar particles. We first note that with this selection, both low and high mass galaxies tend to have low $V/\sigma$, while intermediate mass galaxies have $V/\sigma$ values more consistent with those of a disc-like population, as shown in Figure~\ref{fig:vsig}. In Figure~\ref{fig:dominor}, we show the correlation between spin or minor axis with the separation vector for the three mass bins. The left and middle panels show the minor axis-separation correlation constructed using the simple (left) and reduced (middle) inertia tensor; and the right panel shows the spin-separation correlation.The left panel of Figure~\ref{fig:dominor} shows that there is a significant correlation of the minor axis of high-mass galaxies aligned perpendicular to the separation vector towards other high mass galaxies. In the case of the reduced inertia tensor, the significance of this signal is decreased, consistently with the fact that the reduced inertia tensor yields rounder shapes for these galaxies. For low mass galaxies, the results are similar but with lower significance. For intermediate mass galaxies, we find a small negative correlation between the minor axis and the separation vector for the simple inertia tensor, and a small positive correlation when the reduced inertia tensor is used. This suggests that the reduced inertia tensor minor axis is correlated with the direction of the spin for this disc-like population, while this is not the case for the simple inertia tensor. We interpret this as a decreasing tendency of stellar particles to settle on a disc as a function of distance to the center of mass of the galaxy, possibly tracing merging structures. Indeed, satellite mergers within the \hagn simulation tend to redistribute their angular momentum significantly
within the host (Welker et al., in prep.). 

\begin{figure} 
\centering
\includegraphics[width=0.45\textwidth]{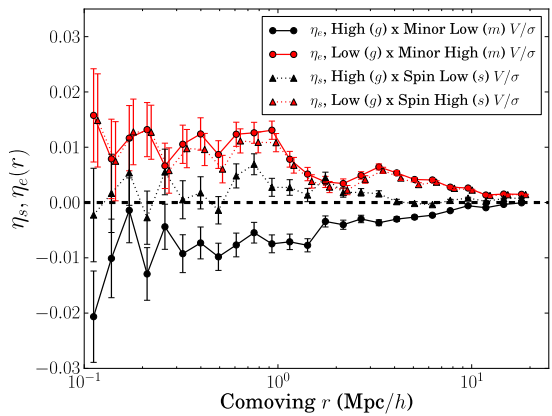}
\caption{Cross-correlations of galaxy positions and shapes ($\eta_e$) or spins ($\eta_s$) for the galaxy population divided in two $V/\sigma$ bins. $\delta$ indicates tracers of the galaxy density field and $s$ ($m$) indicates the sample for which the orientation is measured from the spin (minor axis). The black solid line shows the correlation between the positions of disc-like tracers of the density field ($V/\sigma>0.55$) and the direction of the minor axis of the reduced inertia tensor of spheroidal galaxies ($V/\sigma < 0.55$). The red solid line shows the correlation between the positions of spheroidal galaxies and the direction of the minor axis of disc-like galaxies. The red dotted line shows the correlation between the position of discs and the spin of spheroidals, and the red line shows the correlation between the positions of spheroidals and the direction of the spin axis of discs. (Notice that the red lines almost overlap, showing that the spin and the minor axis of a disc point along the same direction.) These results suggest that the discs are preferentially clustered in the direction of the elongation of spheroidals, while they also tend to have their spins aligned in the direction of nearby spheroidals.}
\label{fig:crossvsig}
\end{figure}

The overall physical picture that we get from these results is the following.
\begin{itemize}
\item Spin and shape alignments depend on galaxy dynamics, and these trends also translate into a mass dependence.
  
\item Spheroidal galaxies show a significant trend of radial alignments with respect to each other that is preserved to large separations and is more prominent for high mass galaxies. The signal decreases in amplitude when the reduced inertia shapes are adopted, as the galaxy shapes become rounder and less sensitive to tidal debris in the outskirts.

\item Spin alignment trends are tangential around other galaxies and marginal, and seem to be better correlated with the reduced inertia tensor than with the simple inertia tensor shapes for intermediate and high $V/\sigma$ galaxies (correspondingly, also intermediate mass galaxies). 
\end{itemize}

\begin{figure}
\centering
\includegraphics[width=0.45\textwidth]{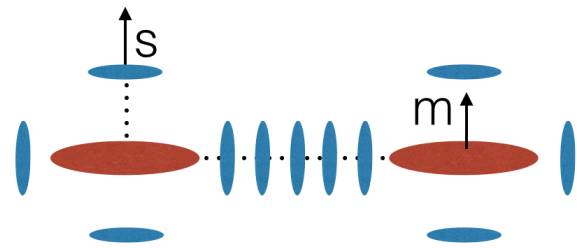}
\caption{A cartoon picture of alignments, as interpreted from the results of Section \ref{sec:align3d}. Discs live in filaments connecting ellipticals and they tend to align their spin({\bf s})/minor axes in the direction of the filament. Ellipticals tend to have their shapes ({\bf m} represents the minor axis) aligned towards each other and towards the direction of the filaments. The effect of alignments is exaggerated for visual purposes by showing all galaxies perfectly aligned following the measured trends in the simulation.}
\label{fig:cartoon}
\end{figure}

The results presented so far only consider correlations between galaxies with similar properties. We also explore whether cross-correlations between subsets exist. Figure~\ref{fig:crossvsig} shows the cross-correlation of positions of spheroidal tracers ($V/\sigma<0.55$) with the direction of the minor axis (red circles) and the spins (red triangles) of disc-like tracers ($V/\sigma > 0.55$). Discs show a tendency to align their spins parallel to the separation vector to spheroidal galaxies (red dashed curve), as well as their minor axis (red solid curve). Notice that the direction of the spin and the minor axis of a disc galaxy (from the reduced inertia tensor) are very well correlated, and hence the red dashed and solid curves almost lie almost on top of each other. Figure~\ref{fig:crossvsig} also shows the cross-correlation of the positions of discs with the orientations of the minor axis (black circles) and the spin (black triangles) of spheroidal galaxies. These galaxies show a preferential elongation of their shapes towards the positions of discs, while they do not show significant alignment of their spins. When the simple inertia tensor shapes are used, the spins and shapes of discs become decorrelated and the significance of the shape alignment trend of discs around spheroidals is lost. From these results, we conclude that the discs are preferentially clustered in the direction of the elongation of spheroidals, while they also tend to have their spins aligned in the direction of nearby spheroidals (also traced by their reduced inertia tensor shapes). This picture is in agreement with disc galaxies living predominantly in a filamentary structure that follows the elongation of spheroidal galaxies at its knots. Furthermore, disc galaxies tend to have their spins aligned parallel to the direction of these filaments, and perpendicular to the elongation of the central spheroidal. Figure \ref{fig:cartoon} shows a cartoon picture of alignments where the effect is exaggerated for visual purposes.

We perform a series of tests to determine the significance of the measured signal of disc alignments with respect to spheroidal tracers. In particular, in this case we adopt a jackknife procedure (similar to that adopted for projected correlations) to estimate the uncertainty in the disc alignment signal. The significance of the measurement is obtained from a $\chi^2$ test using only the diagonals of the jackknife covariance matrix and without modelling the noise in this matrix. We find that the null hypothesis can be rejected at $>99.99\%$ confidence level (C.L.) level for the alignment of disc spins around spheroidals (red dashed in Figure~\ref{fig:crossvsig}), and similarly for the alignment of the minor axes of discs around spheroidals (red solid line in Figure~\ref{fig:crossvsig}). 
This level of significance decreases to $92\%$ when the simple inertia tensor is adopted to measure the shapes. 
On the contrary, the direction of the spin of spheroidals is not correlated with the position of discs ($72\%$ C.L. for null hypothesis rejection of the blue curves),
but the minor axis direction of a spheroidal is anti-correlated with the position of discs with high significance ($>99.99\%$ C.L. for both the reduced ans simple inertia tensor).
Finally, we consider whether alignment signals are still present when the orientation is defined by the direction of the {\it major} axis. We find that the disc alignment measurement in this case is more sensitive to the choice of reduced/simple inertia tensor. This result confirms that the simple inertia tensor is a worse tracer of the spin compared to the reduced inertia tensor, as the alignment signal loses significance in that case.  On the other hand, the use of the simple or reduced inertia tensor does not change the significance of the alignment of spheroidals in the direction of discs.

We conclude that:
\begin{itemize}
\item discs show a significant tendency for tangential alignment around over-densities traced by spheroidal galaxies,
  
\item spheroidals are preferentially elongated towards discs and other spheroidals,

\item and that spin is a good tracer of reduced inertia shapes for discs, but not for spheroidals.
\end{itemize}
  
In the next section, we mimic observations by exploring projected ellipticity alignments.

\subsection{Projected correlations}
\label{sec:resproj}

The intrinsic alignment signal is typically measured in the literature using the projected correlation function of galaxy positions and shapes (Equation~\ref{eq:wgplusdef}). This quantity is readily accessible using shear measurements from survey galaxy catalogs. In this work, we also obtain the projected correlation functions of the density field and projected shapes, $w_{\delta+}$ and $w_{\delta\times}$, where the density field is obtained from a random subsampling of $0.007\%$ of the DM particles in the box. This subsampling guarantees a $10\%$ convergence level in the DM power spectrum, which is similar to the expected level of convergence in determining galaxy shape (as we discussed from Figure~\ref{fig:sigmares}). \citet{Tenneti14b} used a similar approach with a comparable subsampling fraction. The measurement of $w_{\delta+}$ and $w_{\delta\times}$ is advantageous in that it allows us to avoid modelling galaxy bias, or to make any assumption about its scale dependence. Also, we do not need to model peculiar velocities, as the DM and galaxy positions in the box are perfectly known.

\begin{figure*}
\centering
\includegraphics[width=0.45\textwidth]{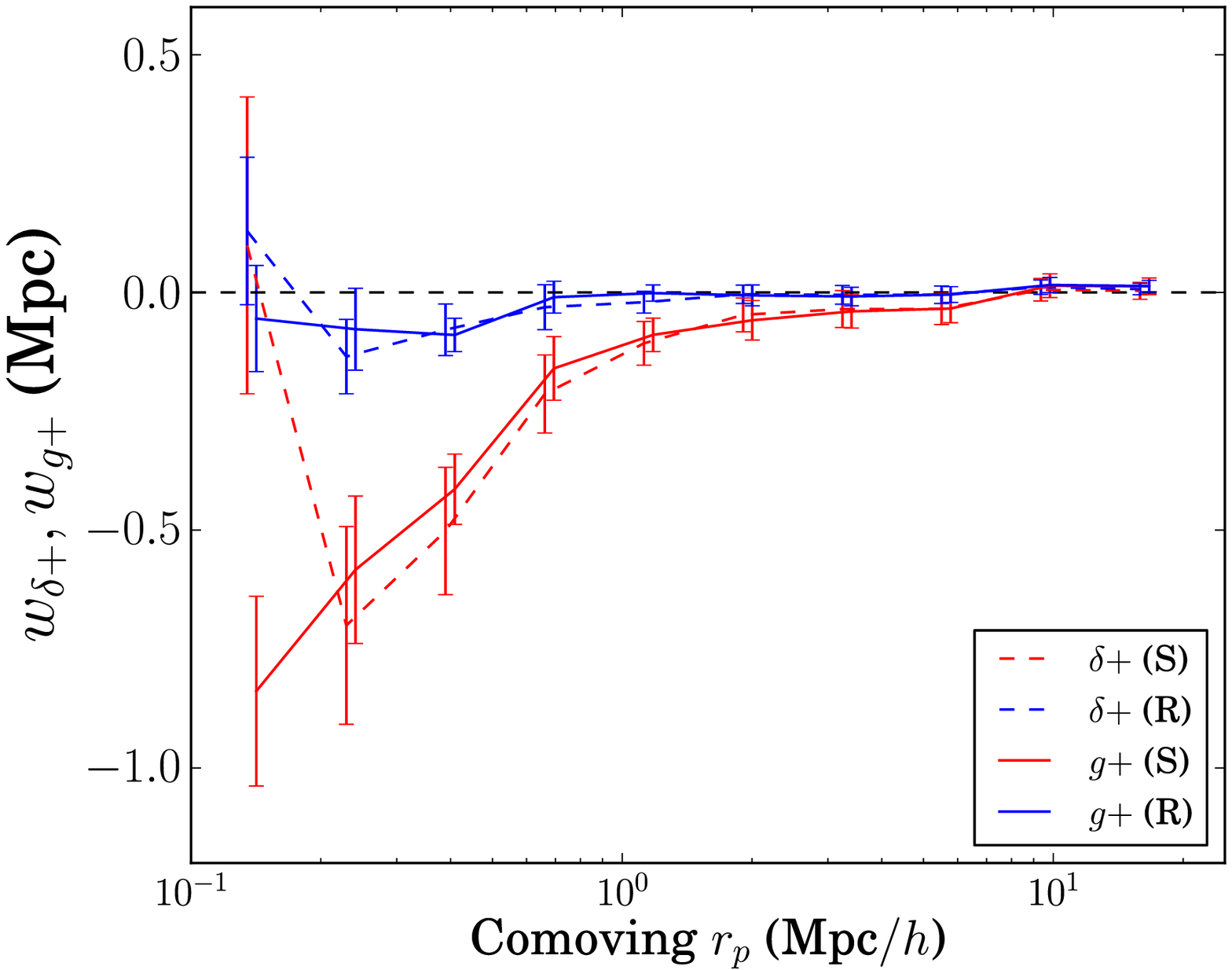}
\includegraphics[width=0.45\textwidth]{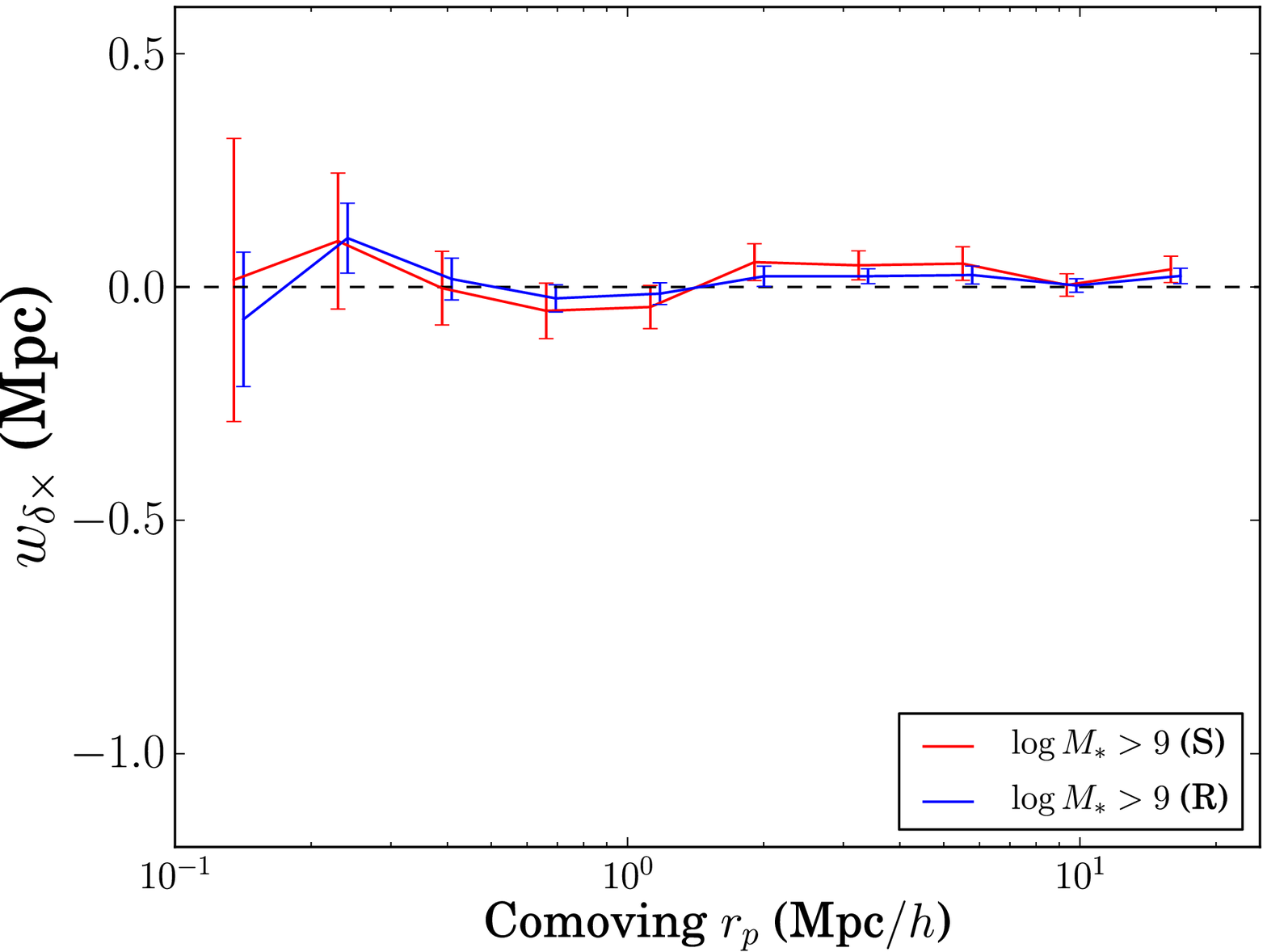}
\caption{$w_{\delta +}$ projected correlation function for all galaxies with $>300$ stellar particles (left panel). This cut corresponds to a cut in mass of $\log (M_*/{\rm M}_\odot) > 9$. The same panel also presents a comparison between the $w_{\delta +}$ correlation and the correlation between galaxy positions and $+$ component of the shapes $w_{g +}$. The right panel shows the projected $\delta\times$ correlation as a test for systematics. In both panels, results obtained with the simple inertia tensor are indicated with the blue line; while the red line corresponds to shapes obtained from the reduced inertia tensor. The measured points for $w_{g+}$ are arbitrarily displaced to larger radii by $5\%$ for visual clarity in the left panel, and similarly for the reduced inertia tensor in the right panel. 
}
\label{fig:wdmshape}
\end{figure*}

We measure the $w_{\delta+}$ correlation function following Equation~(\ref{eq:wgplusdef}) for all galaxies with $>300$ stellar particles in the simulation box and replacing the density tracers by the subsample of DM particles. Grid locking (see Appendix~\ref{sec:gridlock}) is not expected to contaminate the measurement by spherical symmetry. As a consequence, the $S_+R$ term is not expected to contribute to this correlation and we neglect it in this section. Section~\ref{sec:wprojrand} provides confirmation of these assumptions. We show the projected correlation functions of the density field and the $+$ component of the shape from the simple inertia tensor and the reduced inertia tensor in the left panel of Figure~\ref{fig:wdmshape}. As expected from our results in Section~\ref{sec:align3d}, we find an anti-correlation between the $+$ component of the shape and the density field that is significant at $>99.99\%$ C.L. level for the simple inertia tensor. The negative sign indicates that the projected shapes of galaxies are elongated pointing towards other galaxies, i.e., that alignments are radial. We find a decreased tendency for alignments ($47\%$ C.L.) when using the reduced inertia tensor, consistently with rounder shapes and with the results presented in Section \ref{sec:align3d}. The right panel of Figure~\ref{fig:wdmshape} shows that the $\delta\times$ correlation is consistent with null (at the $\simeq 65\%$ C.L.). 

\begin{figure*}
\centering
\includegraphics[width=0.33\textwidth]{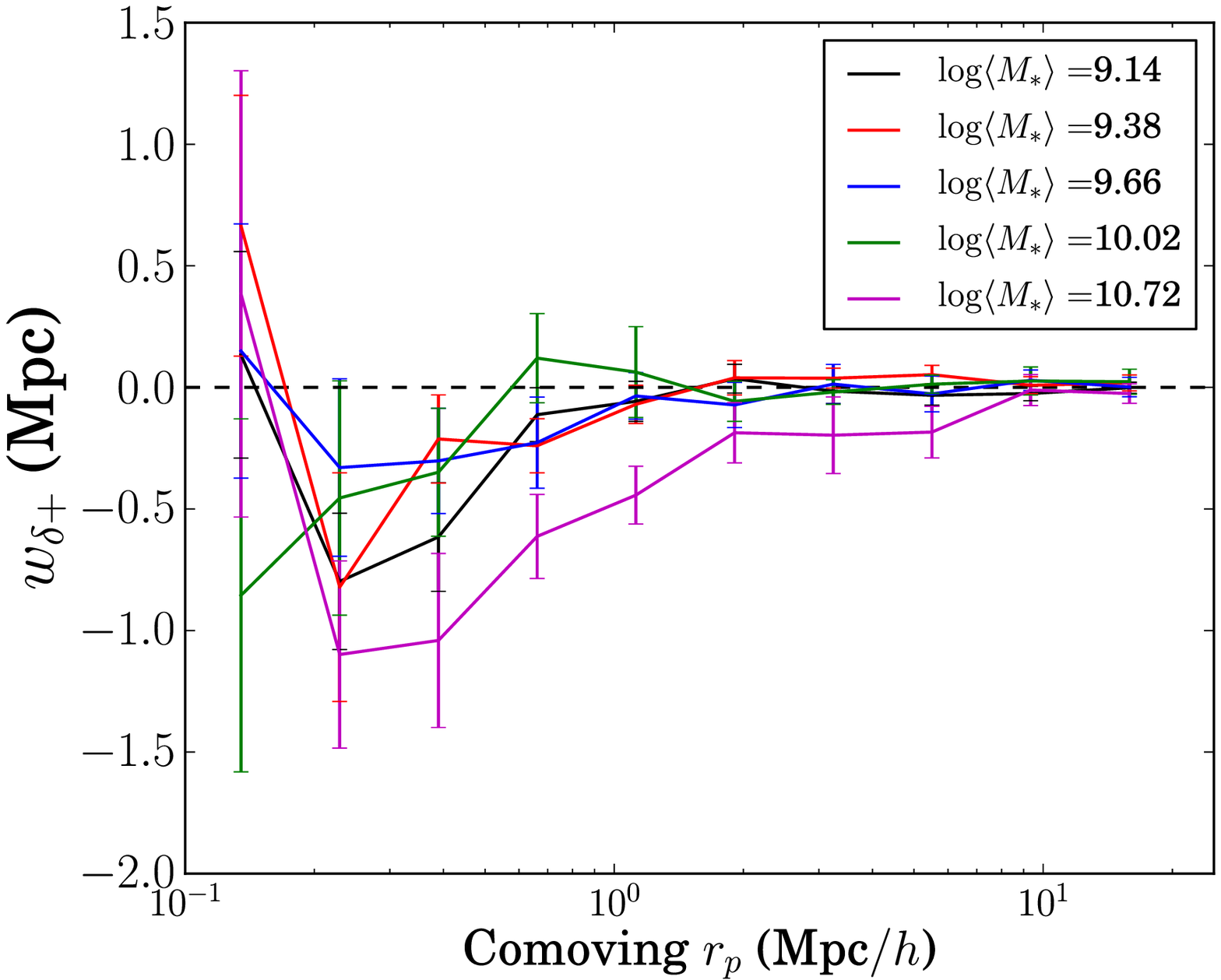}
\includegraphics[width=0.33\textwidth]{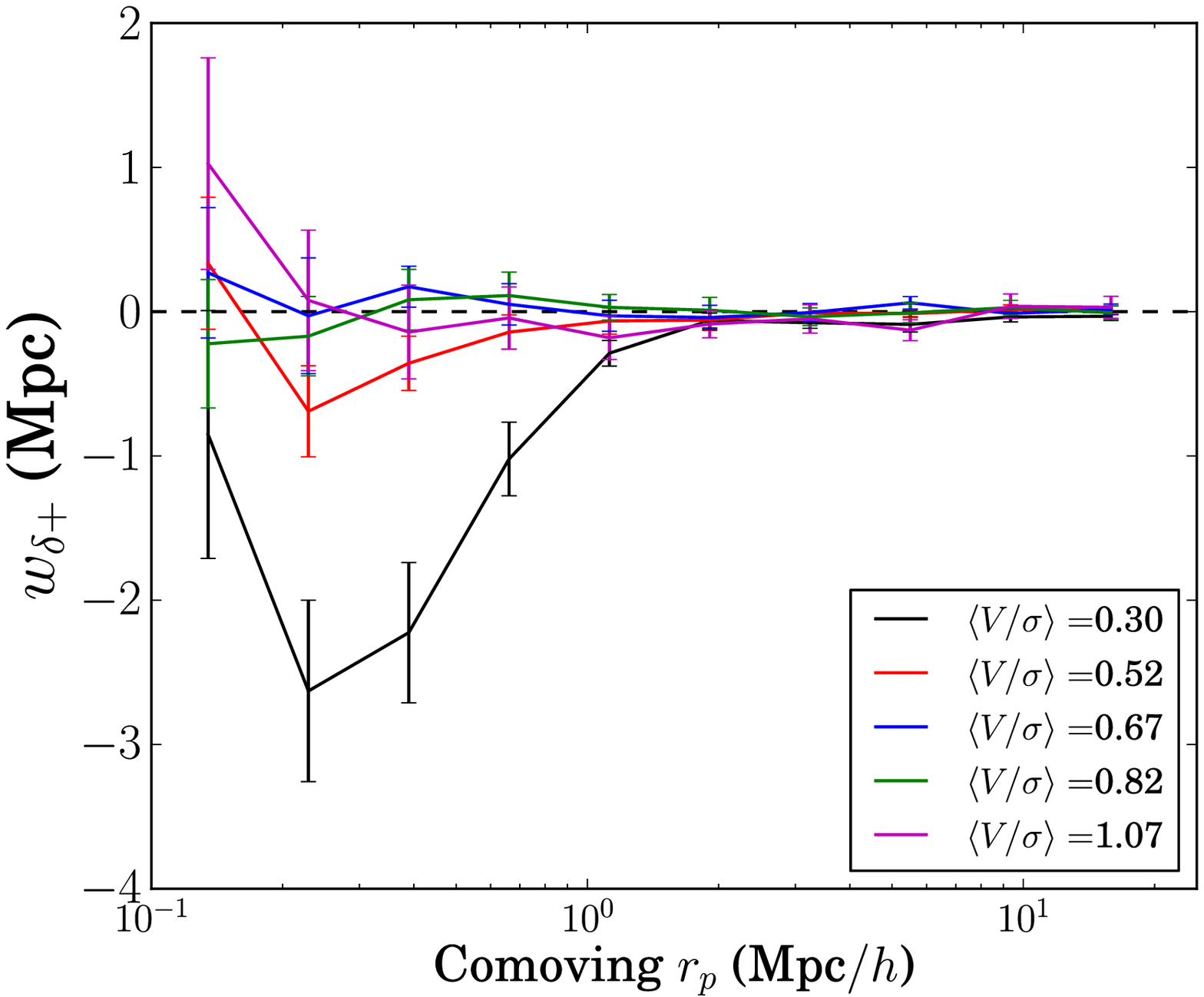}
\includegraphics[width=0.33\textwidth]{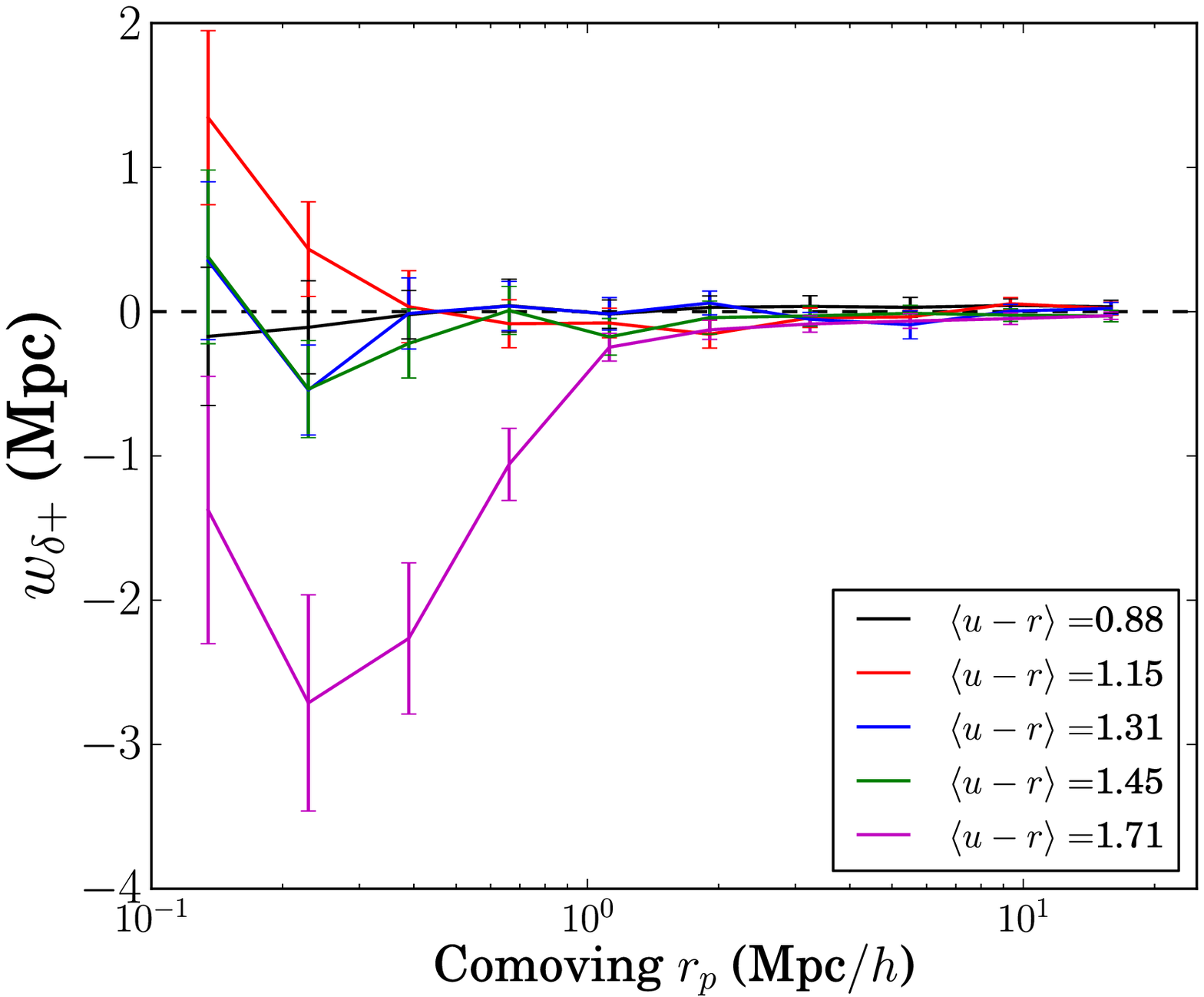}
\caption{$w_{\delta +}$ projected correlation functions for $5$ bins of mass (left panel) and $V/\sigma$ (middle panel) and $u-r$ colour (right panel). The legend indicates the mean of the considered property in each bin, and the $5$ bins are approximately equally populated. For simplicity, we only show the correlations that correspond to shape measurements using the simple inertia tensor. The impact of using the reduced inertia tensor is shown in Figure~\ref{fig:wdmshape}.
}
\label{fig:wdmshape_cuts}
\end{figure*}

\begin{figure}
\centering
\includegraphics[width=0.45\textwidth]{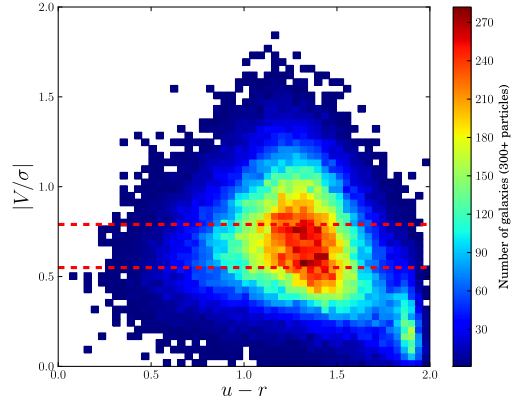}
\caption{Galaxies in the $V/\sigma$ vs $u-r$ plane. There is a strong correlation between $V/\sigma$ and $u-r$ colour at the red end of the colour distribution. Galaxies in \hagn have a bimodal distribution of colours and $V/\sigma$. The red horizontal line represent our fiducial cuts in $V/\sigma$ used in Section~\ref{sec:align3d}.}
\label{fig:vsigur}
\end{figure}

In Figure~\ref{fig:wdmshape_cuts}, we split the sample of galaxies with shapes into $5$ bins of mass (left panel), $V/\sigma$ (middle panel) and $u-r$ colour (right panel). All bins have approximately the same number of galaxies and the legend in each panel indicates the mean of the property considered for the galaxies in each bin. We find that galaxies in the lowest (highest) $V/\sigma$ ($u-r$ colour) bin carry the strongest alignment signal. We find very similar results when splitting the galaxies by their $g-r$ colour. In comparison, the split by mass results in a less clear identification of which galaxies are responsible for projected shape alignments. Figure~\ref{fig:vsigur} shows that there is a significant correlation between colour and $V/\sigma$ for galaxies with redder colours. On the contrary, $V/\sigma$ and mass are not monotonically correlated, as shown earlier in Figure~\ref{fig:vsig}.


\begin{figure}
\centering
\includegraphics[width=0.45\textwidth]{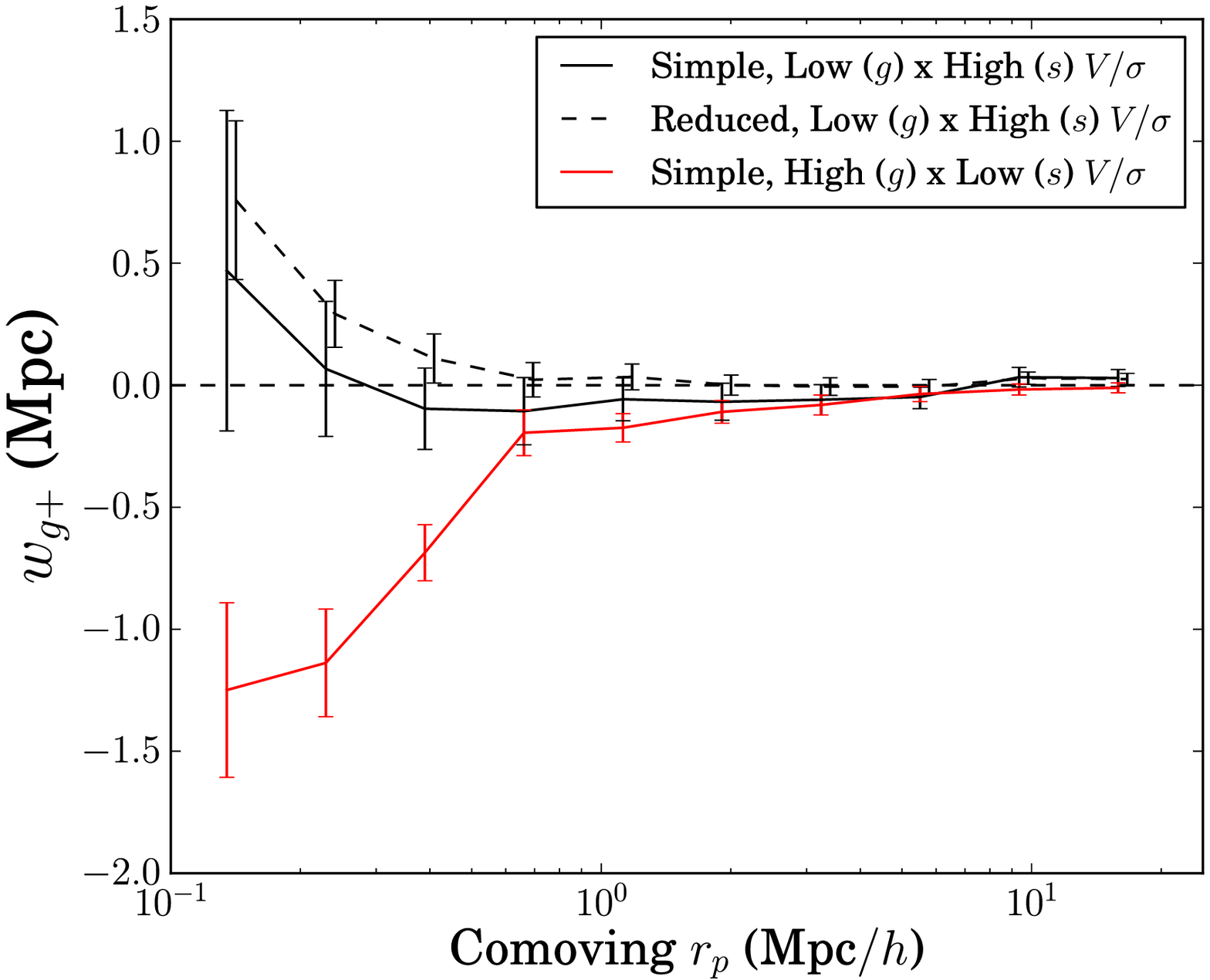}
\caption{Cross-correlation of galaxy positions and $+$ component of the shape applying the same selection as in Figure~\ref{fig:crossvsig}. The alignment of low $V/\sigma$ galaxies in the direction of high $V/\sigma$ galaxies (red) is significant after projection of the shapes and through the simulation box. On the contrary, the tangential alignment of high $V/\sigma$ galaxies around low $V/\sigma$ tracers of the density field (black) is diluted in projection, albeit the dashed line still rejects the null hypothesis at $89\%$ C.L. The error bars in the case of the black dashed line have been artificially displaced to $5\%$ larger $r_p$ for visual clarity.}
\label{fig:projcrossvsig}
\end{figure}

Interestingly, the high $V/\sigma$ galaxies do not show any significant alignment in Figure~\ref{fig:wdmshape_cuts}. This is puzzling given the results presented in Figure~\ref{fig:crossvsig}. To elucidate this discrepancy, we compute the $w_{g+}$ statistic using the same selection cuts as for Figure~\ref{fig:crossvsig} and show the results in Figure~\ref{fig:projcrossvsig}. We find that, while the alignment of spheroids in the direction of the clustering of discs is still significant in projection, the disc alignment signal is diluted and consistent with null at the $27\%$ C.L. using the simple inertia tensor, but less so ($89\%$ C.L.) using the reduced inertia tensor. In the latter case, the signal was more significant from the orientation-separation correlation of Section~\ref{sec:align3d}. In projection, a {\it tangential alignment} is only marginally present. (We remind the reader that gravitational lensing has the same positive sign for correlations measured between pairs of galaxies with large separations along the line of sight).
Notice that, as discussed in Section~\ref{sec:align3d}, a lower level of correlation is expected for the simple inertia tensor given that this is not a good tracer of spin alignment. Moreover, there are several reasons why the disc orientation correlation observed in Figure~\ref{fig:crossvsig} can be diluted in projection. One factor is the weighting by galaxy ellipticity in Equation~(\ref{eq:splusD}). Face-on discs would carry no signal in this statistic. The second reason for dilution is the fact that Equation~(\ref{eq:wgplusdef}) weights the signal in each $\Pi$ bin equally, while the alignment signal is expected to lose correlation as $\Pi$ increases. In comparison, Figure~\ref{fig:crossvsig} showed the level of alignment as a function of three dimensional separation: pairs with large $\Pi$ have large $r$ in that figure and lower correlation. 

Finally, we also study the cross correlation of galaxy positions and galaxy shapes, considering all galaxies in the simulation box. This correlation will include the effect of galaxy bias,  compared to the DM-shape correlation. We show a comparison of cross correlation of $+$ shapes and DM, and of $+$ shapes and galaxy positions in the left panel of Figure~\ref{fig:wdmshape}. We find that the galaxy-shape correlation traces the DM-shape correlation well within the error bars. There is a discrepancy in amplitude of the correlation in the first bin in the case of the simple inertia tensor. The excess power in the galaxy-shape correlation could arise from tidal debris on small scales or from the increased clustering compared to the DM, but we cannot draw firm conclusions from this comparison. In general, the similarity between the galaxy-shape correlation and the DM density-shape correlation suggests that the bias parameter $b_g$ is not very different from unity for the sample of galaxies considered, and that it does not have any significant scale dependence.


\subsection{Modelling of alignment signal}
\label{sec:NLAconstrain}
Early-type galaxy alignments are thought to arise due to the action of the tidal field. In this model, the tidal field contributes a small component to the projected ellipticity of a galaxy, given by \citep{Catelan01}

\begin{equation}
\gamma_{(+,\times)}^I = \frac{C_1}{4\pi G}(\partial_x^2-\partial_y^2,\partial_x\partial_y)\mathcal{S}[\phi_p],
\end{equation}
where $C_1$ is a proportionality constant that parametrizes the response of the shape of a galaxy to the tidal field, $\phi_p$ is the Newtonian gravitational potential at the redshift of formation of a galaxy and $\mathcal{S}$ is a smoothing filter that acts to smooth the potential over the typical scale of the galactic halo ($\sim 1$ Mpc)\footnote{Notice that we adopt a different sign convention than \citet{Singh14} and \citet{Tenneti14b}, whereby alignments are negative if they are radial, and positive if tangential.}. As a consequence, there is a correlation between galaxy positions and their intrinsic shapes, given by

\begin{equation}
  P_{g+}({\bf k},z) = -\frac{b_gC_1\rho_{\rm crit}\Omega_m}{D(z)}\frac{k_x^2-k_y^2}{k^2}P_\delta({\bf k},z)\,,
  \label{eq:powerdi}
\end{equation}
where $b_g$ is the galaxy bias, $\rho_{\rm crit}$ is the critical density of the Universe today, $D(z)$ is the growth function (normalized to unity at $z=0$) and $P_\delta$ is the matter power spectrum. While these expressions are not strictly valid in the non-linear regime, it is customary to approximate the nonlinear scale alignments by replacing the linear power spectrum in Equation (\ref{eq:powerdi}) \citep{Bridle07} by its nonlinear analogue. Recent observational works are beginning to test this assumption \citep{Singh14}, but most constraints on the amplitude of alignments are still typically given in terms of the nonlinear alignment (`NLA') approximation of \citet{Bridle07}. The $g\times$ power spectrum is not presented because it is expected to average to null in projection.

Equation (\ref{eq:powerdi}) can be transformed to redshift space to give a prediction for the on-the-sky $w_{g+}$ projected correlation,

\begin{align}
w_{g+}(r_p) &= -\frac{b_gC_1\rho_{\rm crit}\Omega_m }{\pi^2D(z)} \int_0^{\infty}dk_z \int_0^{\infty}dk_{\perp} \nonumber
\\
& \hskip -1cm \frac{k_\perp^3}{(k_{\perp}^2+k_z^2)k_z}P_\delta({\bf k},z)\sin(k_z\pi_{\rm max})J_2(k_{\perp} r_p),
\label{eq:wplus_nla}
  \end{align}
where $b_g$ is the galaxy bias, $r_p$ is the projected radius, $\Pi_{\rm max}$ is the line of sight distance over which the projection is carried out and $J_2$ is the second order Bessel function of the first kind. Notice that the Kaiser factor \citep{Kaiser87,Singh14} is unnecessary to model the galaxies in the simulation, since we have access to the true positions of the galaxies along the line of sight. Similarly, we can safely neglect fingers-of-God effects arising from peculiar velocities in the nonlinear regime. 

\citet{Singh14} observed an excess of power on small scale alignments compared to the best fit NLA model to the LOWZ galaxy sample of the SDSS. In this regime, we also consider a halo model developed by \citet{Schneider10}. According to this model, the one halo power spectrum of galaxy positions and intrinsic shapes is given by
\begin{equation}
P_{\delta,\gamma^I}^{1h}(k,z)=-a_h\frac{(k/p_1)^2}{1+(k/p_2)^{p_3}}\,,
\end{equation}
where $a_h$ is the halo model alignment amplitude on small scales; $p_1$, $p_2$, and $p_3$ are fixed parameters based on fits by \citet{Schneider10}; and the projected correlation function is then given by
\begin{equation}
w_{g+}^{1h}= -b_g \int \frac{dk_\perp}{2\pi}\,k_\perp P_{\delta,\gamma^I}^{1h}(k_\perp,z)J_0(k_\perp r_p).
\label{eq:oneh}
\end{equation}

While the alignments of disc galaxy shapes have not been detected in observations, and despite the fact that the tidal alignment model is not expected to describe their alignments \citep{Catelan01,Hirata04}, the available constraints are typically phrased in terms of the NLA model as well \citep[e.g.,][]{WiggleZ}.

We now place quantitative constraints on the goodness-of-fit of the intrinsic alignment models. We build a model template of the intrinsic alignment signal, $w_{\delta+}^\mu(r_p)$, at the positions for which we measure this correlation in the simulation. We model the nonlinear matter power spectrum using the CAMB software \citep{Howlett12} with the HALOFIT
correction \citep{Smith03}.

Using the diagonals of the jackknife covariance, we determine the $\chi^2$ from summing over all radial bins as follows
\begin{equation}
\chi^2 = \sum_{r_p}\frac{(w_{\delta+}(r_p)-w_{\delta+}^\mu(r_p))^2}{{\rm Var}[w_{\delta+}(r_p)]}\,,
\end{equation}
where $\mu$ represents the alignment model template to be fit and we look for the minimum $\chi^2$ by varying the parameters of the fit; and analogously for $w_{g+}$, in which case only joint constraints of the product of $b_g$ and the alignment amplitude are obtained. We emphasize that we are not modelling covariances or noise in this matrix when performing the fits.

Notice that while the results presented in the previous section do not guarantee that disc-like galaxies do not contribute to the alignment signal, the correlation functions shown in Section \ref{sec:align3d} suggest that they counter-act the radial alignment of ellipticals due to their tendency to orient tangentially around over-densities. For this reason, and given that we do not have sufficient constraining power to bin the galaxy sample into different populations, we fit different models to $w_{g+}$ and $w_{\delta +}$ shown in Figure~\ref{fig:wdmshape}, which include the contribution of all galaxies with $>10^9 {\rm M}_\odot$ stellar masses in the simulation.

\begin{figure*}
\centering
\includegraphics[width=0.7\textwidth]{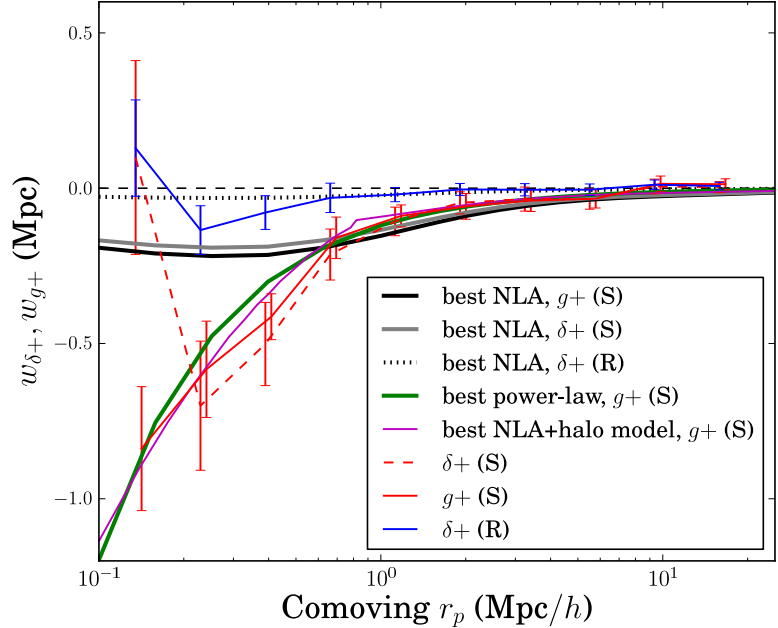}
\caption{Alignment model fits for $w_{\delta +}$ and $w_{g+}$ for all galaxies with $>300$ stellar particles. Results obtained with the simple inertia tensor are indicated with the red lines; while the blue line corresponds to shapes obtained from the reduced inertia tensor for $w_{\delta+}$ only. We show fits from the NLA model to $w_{\delta +}$ (gray solid for simple inertia tensor; black dotted for reduced inertia tensor) and to $w_{g+}$ (black solid for the simple inertia tensor). The measured points for $w_{\delta+}$ and simple inertia tensor shapes are arbitrarily displaced to larger radii by $5\%$ for visual clarity. The NLA model significantly underestimates the power in alignments at small separations ($< 1$ Mpc$/h$) in qualitative agreement with observations by \citet{Singh14}. 
}
\label{fig:wdmshape_wfits}
\end{figure*}

We focus on fits to the signal obtained using the simple inertia tensor, as in the case of the reduced inertia tensor the results are consistent with null. A power-law fit as a function of projected radius ($w^\mu_{g+}=b_gA_Ir_p^{\beta}$) yields the following constraints on power-law amplitude and power-law index: $b_gA_I=-0.13\pm0.03$ and $\beta=-0.75\pm0.25$, respectively. The error bars quoted correspond to $68\%$ C.L when holding one of the parameters fixed at the best fit value corresponding to the minimum $\chi^2$ ($b_gA_I=-0.17$ and $\beta=-1.0$).
We find that the linear alignment model is a very poor fit due to the lack of power on small scales compared to the measured signal. Similarly, the NLA model also underestimates the correlation of positions and shapes on small scales\footnote{In this particular work, we do not apply any  smoothing to the tidal field. A smoothing filter would suppress power on small scales, worsening the comparison of the NLA model to the measured signal.}. In this case, the best fit model is shown in Figure~\ref{fig:wdmshape_wfits}. The best fit NLA model for $w_{g+}$ in the case of the simple inertia tensor is represented by the black solid line. In gray, we show the best fit NLA model for $w_{\delta+}$, also for the simple inertia tensor, which is comparable to the black line. The black dotted line shows the best fit to $w_{\delta+}$ in the case of the reduced inertia tensor, for which the best fit amplitude is consistent with null at $68\%$ C.L.

Given that the best fit NLA model underestimates the alignment signal on small scales, we consider fitting a sum of the NLA model template (across all scales measured) and the halo model (only at $r_p<0.8/h$) with the parameters of \citet{Schneider10}. Notice that the physical interpretation of these results is not straightforward and should only be considered as phenomenological. The constraints on the parameters for the NLA and halo model combination are: $b_ga_h=0.27^{+0.12}_{-0.11}$, $b_gA_I=0.58^{+0.34}_{-0.35}$. The $\chi^2$ in this case is comparable to that of the power-law fit. If we use more and smaller ($L/4$ a side) jackknife regions, the error bars increase slightly yielding $b_gA_I=0.60_{-0.41}^{+0.51}$.

Given the limited simulation volume and the corresponding (cosmic) large error bars, we cannot obtain meaningful constraints on mass or luminosity dependence of the signal. However, we note that, as discussed in Section \ref{sec:resproj}, the mass dependence is not monotonic. 

\section{Discussion}
\label{sec:discuss}

Theoretical models of intrinsic alignments have indeed suggested that the population of galaxies subject to this mechanism can be split in two \citep{Catelan01}. Disc-like systems interact with the large-scale structure through torques to their angular momentum vector, while spheroidals, which do not have significant angular momentum, would tend to orient their major axes pointing towards over-densities. In this work, we have used $V/\sigma$ as a proxy of galaxy morphology, and we have indeed confirmed the existence of two different alignment mechanisms in play for spheroidals (low $V/\sigma$) and disc-like (high $V/\sigma$) galaxies. While $V/\sigma$ is not easily accessible to upcoming gravitational lensing surveys, fortunately there is a strong correlation between $u-r$ and $V/\sigma$ (Figure~\ref{fig:vsigur}) that can help identify the two populations of galaxies and their alignments. Low $V/\sigma$ galaxies tend to have redder colours, while high $V/\sigma$ galaxies tend to be bluer. 

The alignments of blue galaxy shapes have so far been consistent with null from observations \citep{WiggleZ,Heymans13}, albeit with large error bars that still allow for a significant level of contamination from blue galaxy alignments to cosmological observables in current and future surveys \citep{Chisari15,Krause15}. We similarly find that the projected correlation function of blue galaxy positions and shapes is consistent with null in \hagn\!\!\!. Nevertheless, while blue galaxies do not align around each other, they tend to align tangentially around red galaxies in three dimensional space, and this signal is washed out in projection {\it within our error bars}. We emphasize that this does not imply that blue galaxy alignments can be neglected for future surveys. \citet{Codis14} found that spin-spin alignments could potentially translate into worrisome levels of contamination for future lensing surveys, particularly for blue galaxies. Their results were based on a ``spin-gives-ellipticity'' prescription for translating spins into shapes for disc-like galaxies. We found that the validity of this assumption depends on the method used to determine ellipticity, with better agreement for the reduced inertia tensor case. The statistics of spin/shape alignments of disc galaxies in three dimensions from the simulation can be a useful tool to constrain disc alignment models and determine the level of contamination to future surveys. 

Several theoretical predictions have been made for the spin alignments of disc-like galaxies. In simulations, discs tend to form with their spin correlated with the direction of filaments \citep{bailin&steinmetz05,Aragon07,Hahn07,sousbie08,Zhang09,codisetal12,Libeskind12,dubois14} and the vorticity of the density field around them \citep{Laigle15}. This is consistent with the scenario of spin acquisition by tidal torquing biased by the large scale structure filaments. Extended tidal torque theory predicts that halos with mass below $5\times 10^{12}\, \rm M_{\odot}$ at redshift $z=0$ tend to form with their spin pointing along the direction of filaments, the center of which are represented by saddle points of the density field \citep{ATTT}. In the plane perpendicular to the filament and containing the saddle point, halos are predicted to show this preferential orientation. Away from this plane and closer to the nodes of the filament where higher mass halos reside, spins flip direction, becoming perpendicular to the filamentary axis.  Hence, if galactic spins are correlated with halo spins, we should expect low-mass galaxies to have a spin aligned with the filamentary axis (as found in \cite{dubois14} for \hagn\!\!\! galaxies) and therefore also aligned\footnote{or anti aligned, should the galaxies belong to octants of opposite  polarity, see \citet{ATTT}, section 6.2.3.} with the separation vector (since we expect a higher number of galaxies to inhabit within the filament). In contrast, \cite{ATTT} predict that very massive galaxies should have a spin perpendicular to the filament. However, massive galaxies tend to be supported by random motions of their stars, and hence the definition of the spin direction becomes more noisy. The signal predicted by theory could also be diluted due to the fact that not all low mass galaxies reside in filaments. We found that the spin and the minor axes of massive galaxies have less tendency to be aligned than for their lower mass counterparts. 

Works by \citet{Tenneti14a,Tenneti14b,Velliscig15} have also studied the intrinsic alignments of galaxies in hydrodynamical simulations. \citet{Tenneti14b} find that orientation-position correlations (their `ED' correlation) with the density field have similar strengths and signs for blue and red galaxies when correlated with the density field. On the contrary, we find that blue galaxies tend to have {\it tangential alignments} around the locations of red galaxies (but not around each other). In projection, both works find the blue galaxy alignments to be suppressed. Those authors did not study spin alignments, which in this work have been shown to be significant for disc-like galaxies and to be connected to their shape alignments.

\begin{figure}
\centering
\includegraphics[width=0.45\textwidth]{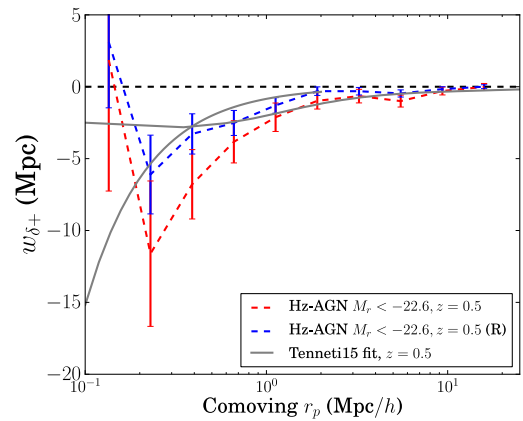}
\caption{Measurement of $w_{\delta+}$ from \hagn for the most luminous galaxies, with absolute $r$-band magnitudes $M_r<-22.6$ ($\sim 800$ galaxies). The blue curve corresponds to shapes measured using the reduced inertia tensor and the red curve, using the simple inertia tensor. We also show alignment model fits from \citet{Tenneti14b} (gray), which are in good agreement with our results. \citet{Tenneti14b} fit a power-law on small scales ($0.1-1$ Mpc$/h$, shown here in the range $0.1-2$ Mpc$/h$) and the NLA model on large scales ($6-25$ Mpc$/h$, extrapolated here over all scales shown). While the agreement is good at high luminosity, we find that the fits by \citet{Tenneti14b} significantly overpredict the alignment signal for the whole sample shown in Figure~\ref{fig:wdmshape_wfits}.
}
\label{fig:tenneti}
\end{figure}

Observational results have shown there is a strong trend for radial alignment of the ellipticities of red galaxies \citep{Mandelbaum06,Hirata07,Okumura09,Joachimi11,Heymans13,Singh14}, which increases for higher mass galaxies. This trend is also seen in \hagn for the shapes of spheroidals. \citet{Tenneti14b} presented NLA and power-law fits to $w_{\delta+}$ from their simulation as a function of redshift and luminosity of the galaxy sample using iterative reduced inertia tensor for the galaxy shapes. We find that their fits reproduce the alignment signal of luminous galaxies ($M_r<-22.6$) presented in this work. We show this agreement in Figure \ref{fig:tenneti}. The gray curves show their power-law fits on small scales, and their NLA model fits across all scales (see their Table 1). Given that \citet{Tenneti14b} match the observed alignments of LRGs with their simulation data, it is expected that \hagn will equally match observations if the redshift dependence of the signal is similar to that found in that work. However, we find that their fits significantly overestimate the alignment amplitude of the whole galaxy sample presented in Figure~\ref{fig:wdmshape_wfits}, and this could indicate a steeper luminosity dependence of alignments in \hagn\!\!\!.  \citet{Joachimi11} indeed found a steeper luminosity dependence of observed LRG alignments compared to \citet{Tenneti14b}. However, such a steep luminosity dependence is not sufficient to reproduce the alignment amplitude of Figure~\ref{fig:wdmshape_wfits}. It is possible that blue galaxy alignments are suppressing our results in that figure as well.

\citet{Codis15} found no alignments for red galaxies using \hagn at $z=1.2$. In that work, spin was used as a proxy for galaxy shape.  Our results are consistent with theirs. We have shown that for red galaxies, the spin alignment signal is very different from the shape alignment signal, as these galaxies do not carry significant angular momentum. 

\citet{Velliscig15} studied the galaxy-halo misalignment comparing the shapes of stars and hot gas to that of the underlying DM halo using the EAGLE smoothed-particle-hydrodynamics simulation. They found that the alignment of the stellar component with the entire DM halo increases as a function of distance from the center to the subhalo and as a function of halo mass. This is in qualitative agreement with our finding that the shape alignment signal is reduced when using the reduced inertia tensor, which puts more weights towards the inner regions of galaxy. However, they also find that misalignment angles between the stellar and the DM components are larger for early-type than for late-type galaxies using the simple inertia tensor. This result will require further comparison, as the DM is expected to have stronger alignments than the baryons \citep{Okumura09} and given that we find a stronger shape alignment signal in \hagn for early-types than for late-types.

\section{Conclusions}
\label{sec:conclusion}

We have studied the alignments of galaxies, as traced by their stellar particles, using the \hagn simulation.

The main result of this paper is the clear identification of two different alignment mechanisms for disc-like galaxies and spheroidals, in qualitative agreement with theoretical expectations \citep{Catelan01,ATTT}. This is the first time that these two mechanisms are clearly separated in a hydrodynamical cosmological simulation. In contrast, previous work by \citet{Tenneti14a} was unable to distinguish between red/blue galaxy orientation-separation alignments. This is likely a consequence of the different methods used to solve for the hydrodynamics (AMR in \hagn compared to smoothed particle hydrodynamics in MassiveBlack II), which result in different galaxy properties and their evolution with redshift.

We also reached the following conclusions:
\begin{itemize}
\item High mass (low $V/\sigma$) galaxies are elongated pointing towards other galaxies. This trend is preserved when projected correlations of the density field and galaxy shapes are considered. 
\item There is a preferential tangential orientation of disc-like galaxies around spheroidals. This trend is diluted in projection, possibly due to the equal weighting of the different $\Pi$ bins in Equation~(\ref{eq:wgplusdef}) and the reduced contribution of face-on discs to Equation~(\ref{eq:splusD}). This suggests that in order to extract the maximum possible information from intrinsic alignments in simulations, it could be beneficial to avoid performing projections along the line of sight, and rather access the full three dimensional information on alignments provided by the simulations. This would increase the signal-to-noise ratio in intrinsic alignments constraints from simulations and allow for more accurate forecasts of intrinsic alignment contamination to weak lensing in future surveys, such as {\it Euclid\,}\footnote{\url{http://sci.esa.int/euclid}}\!~\citep{Laureijs11}, the Large Synoptic Survey Telescope\footnote{\url{http://www.lsst.org}}\citep{Ivezic08} and WFIRST\footnote{\url{http://wfirst.gsfc.nasa.gov/}}~\citep{green11}.
\item We are able to describe $w_{g+}$ across all scales probed and for all galaxies in the simulation with $>10^9 \, \rm M_\odot$ using a power-law in the case where shapes are obtained with the simple inertia tensor. The NLA and LA model tend to underestimate the power on small scales; a conclusion also reached by \citet{Singh14} using low redshift observations of LRGs. Fits to $w_{g+}$ in the reduced inertia tensor case result in an alignment amplitude that is consistent with null for our complete sample of galaxies. This does not imply that alignments are not potential contaminants to weak lensing measurements. It will be necessary to match the shape measurement and galaxy selection done in observations to make more quantitative assessments of contamination from alignment to future surveys. The alignments of the most luminous galaxies in \hagn are, in fact, in agreement with work by \citet{Tenneti14b} and the alignments of bright LRGs in SDSS.
\item Galactic kinematics, as quantified by $V/\sigma$, is a good proxy for the level of alignment. We also find that,  given the existing correlation between $V/\sigma$ and $u-r$ colour, the latter can also be used as proxy to separate galaxy populations with different sensitivity to alignment. On the contrary, the amplitude of the intrinsic alignment signal is not monotonic with stellar mass, which we interpret as a consequence of the wide distribution of stellar masses at low $V/\sigma$.
\item We emphasise that correlations of spins (or shapes) and separation are not contaminated by grid-locking because this effect averages to null for position-shape correlations (as shown in Appendix~\ref{sec:gridlock}). This is not necessarily true for two-point auto-correlations of spins and/or shapes, and so we refrain from giving these a physical interpretation in the main body of the manuscript, although we present them in Appendix~\ref{sec:auto} for completeness.
\item The choice of shape estimator can have a large impact on predictions for intrinsic alignments from cosmological simulations. To make accurate predictions of intrinsic alignment impact on weak lensing surveys, we expect that it will be necessary to create mock images of simulated galaxies in a manner that takes into account photometric depth, noise and convolution by the point-spread function. We defer this and a study of redshift evolution and selection cuts on the galaxy sample to future work.
\end{itemize}

\section*{Acknowledgments}
This work has made use  of the HPC resources of CINES (Jade and Occigen supercomputer) under the time allocations 2013047012, 2014047012 and 2015047012 made by GENCI. This work is partially supported by the Spin(e) grants {ANR-13-BS05-0005} (\url{http://cosmicorigin.org}) of the French {\sl Agence Nationale de la Recherche} and by the ILP LABEX (under reference ANR-10-LABX-63 and ANR-11-IDEX-0004-02). Part of the analysis of the simulation was performed on the DiRAC facility jointly funded by STFC, BIS and the University of Oxford. NEC is supported by a Beecroft Postdoctoral Research Fellowship. We thank  S. Rouberol for running  smoothly the {\tt Horizon} cluster for us. NEC thanks Rachel Mandelbaum for useful discussion regarding the grid-locking effect.

\bibliographystyle{mn2e}
\bibliography{author}

\appendix

\section{Grid-locking }
\label{sec:gridlock}

Adaptive-mesh-refinement simulations are thought to be subject to grid-locking systematics, whereby the galaxy spins become aligned with the directions of the grid. We have verified that this effect is indeed present in the \hagn simulation, but we demonstrate in this appendix that it does not affect position-shape or position-spin correlations.

\subsection{Correlations with the box}
\label{sec:syst}

\begin{figure*}
  \centering
    \includegraphics[width=0.45\textwidth]{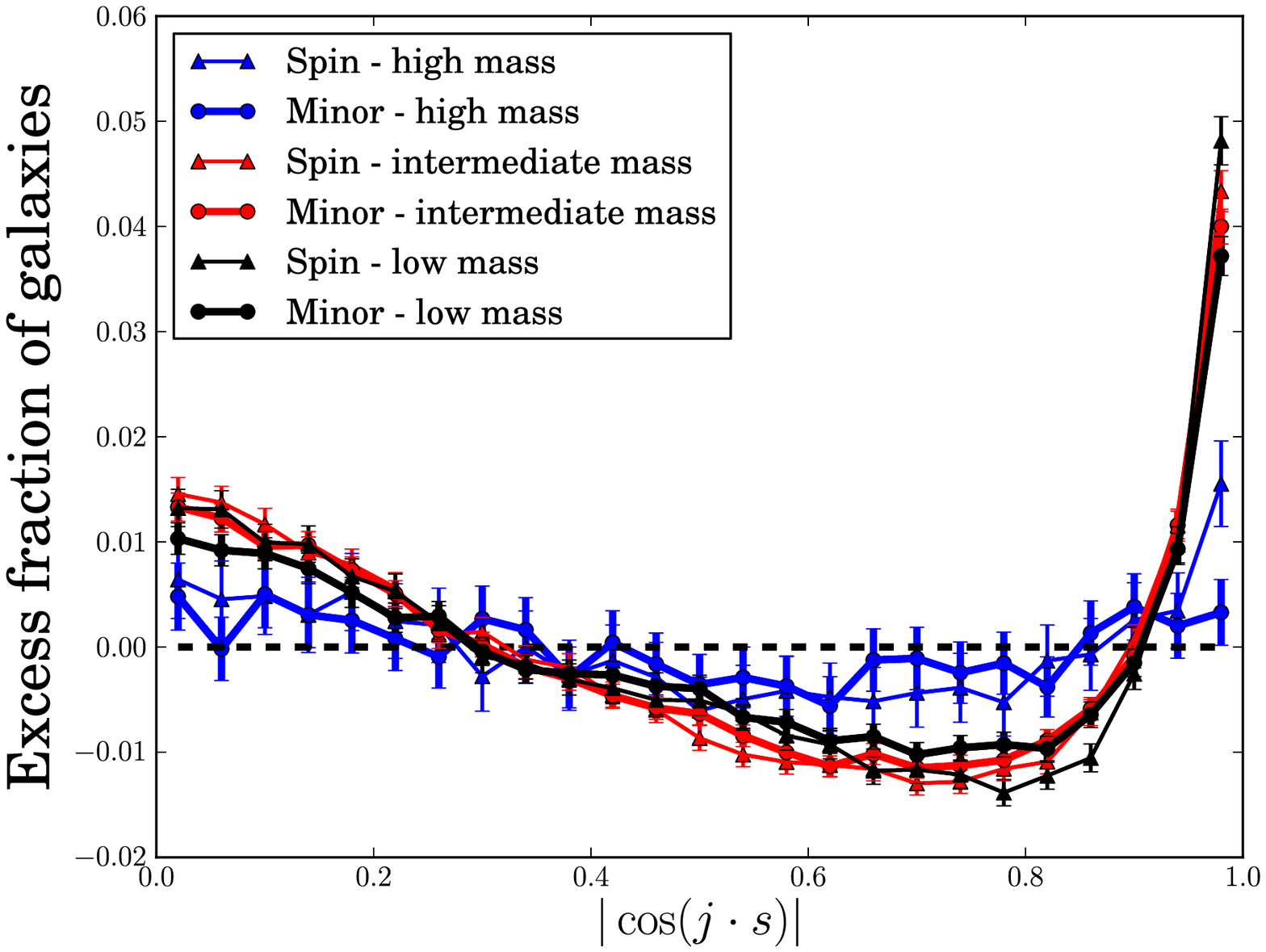}
    \includegraphics[width=0.45\textwidth]{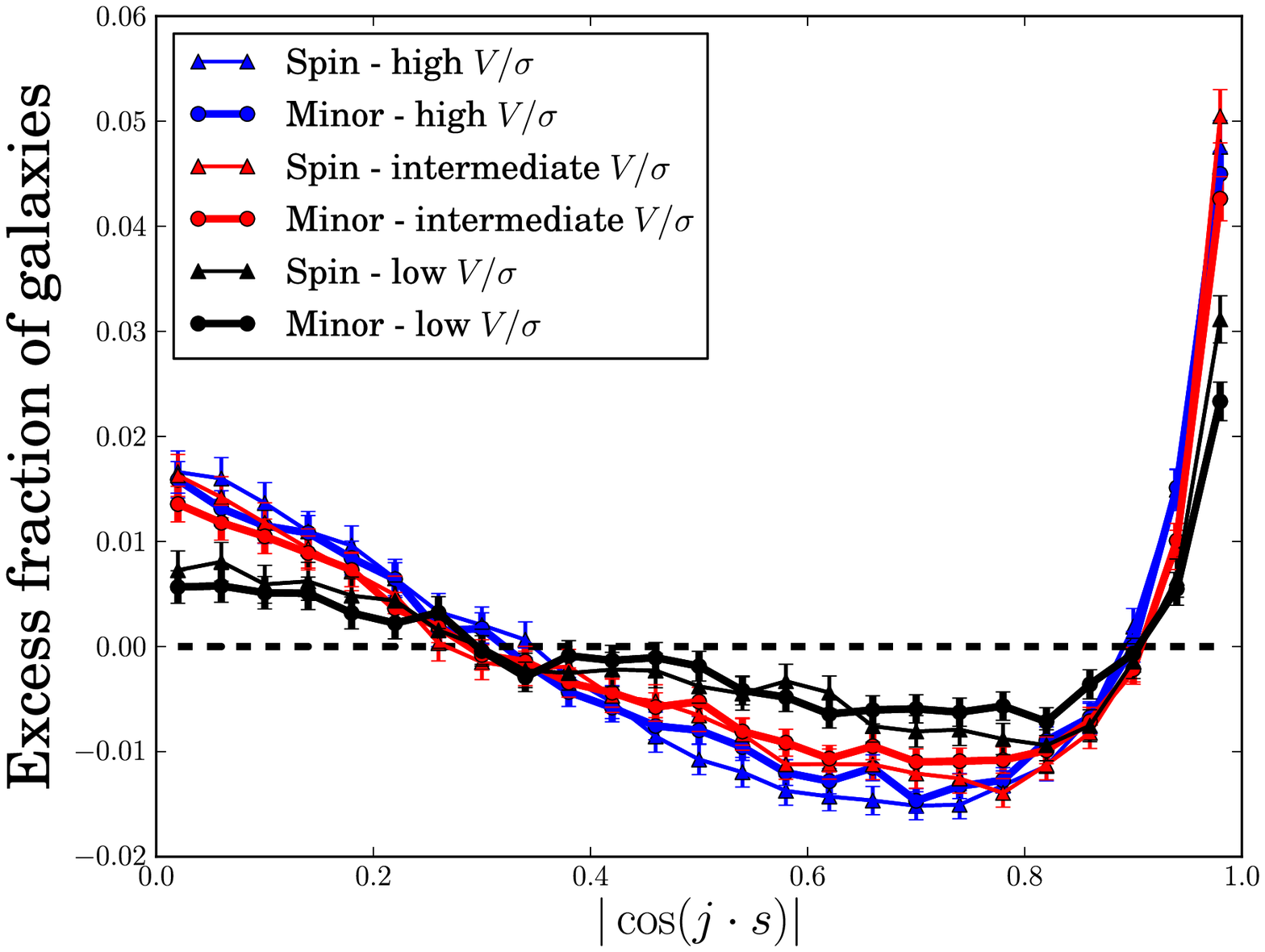}
  \caption{Excess fraction of galaxies with respect to a uniform distribution as a function of the angle between the spin/minor axis with the directions of the box (averaged over $j=\{x,y,z\}$). In the left panels, the galaxy sample is split by mass in the following bins: $10^9\,{\rm M}_\odot<M_*<10^{9.5}\,{\rm M}_\odot$ (black), $10^{9.5}\,{\rm M}_\odot<M_*<10^{10.6}\,{\rm M}_\odot$ (red) and $M_*>10^{10.6}\,{\rm M}_\odot$ (blue). In the left panel, the galaxy sample is split by $V/\sigma$: $V/\sigma<0.55$ (black), $0.55<V/\sigma<0.79$ (red) and $V/\sigma>0.79$ (blue). There is an excess of galaxies in directions parallel and perpendicular to the grid axes, which is less significant for high mass and low $V/\sigma$ galaxies. This effect averages to null for the angular and projected correlations of spin/minor axis and separation, but it could give rise to a non-negligible two-point correlation. 
}
\label{fig:box}
\end{figure*}

We consider possible systematics that might arise from correlations of the galaxy orientations with preferential directions of the simulation box. Figure~\ref{fig:box} shows the excess fraction of galaxies as a function of $|\cos({\bf s}\cdot\{{\bf x,y,z}\})|$, where ${\bf s}$ is the galaxy spin and we average over ${x,y,z}$. There is a clear excess of galaxies at directions perpendicular and parallel/anti-parallel to the grid, and a decrement at intermediate angles. A similar behaviour is observed for the direction minor axes in the same figure. The higher mass galaxies are less affected, but the trend is not monotonic with mass. We have also considered selecting the galaxy sample by median stellar age, resolution and colour. While mass and median stellar age cuts can help remove the systematics, it is not possible to define clean sample without removing the vast majority of the galaxies. Moreover, it is not clear how these selection cuts affect the intrinsic alignment measurement and its comparison to current observational constraints. For these reasons, we decide to avoid placing cuts on the galaxy sample to reduce the grid-locking effects. Instead, we show in the next section that the correlation between galaxy shapes, spins and the box have no impact on the $w_{\delta +}$ statistic presented in Section \ref{sec:correl}.

We have also considered the grid-locking signal of shapes obtained through the reduced inertia tensor. In this case, we find that the grid-locking signal is smaller than for Figure~\ref{fig:box}. 

\subsection{Contamination to intrinsic alignments}
\label{sec:wprojrand}

\begin{figure}
\centering
\includegraphics[width=0.4\textwidth]{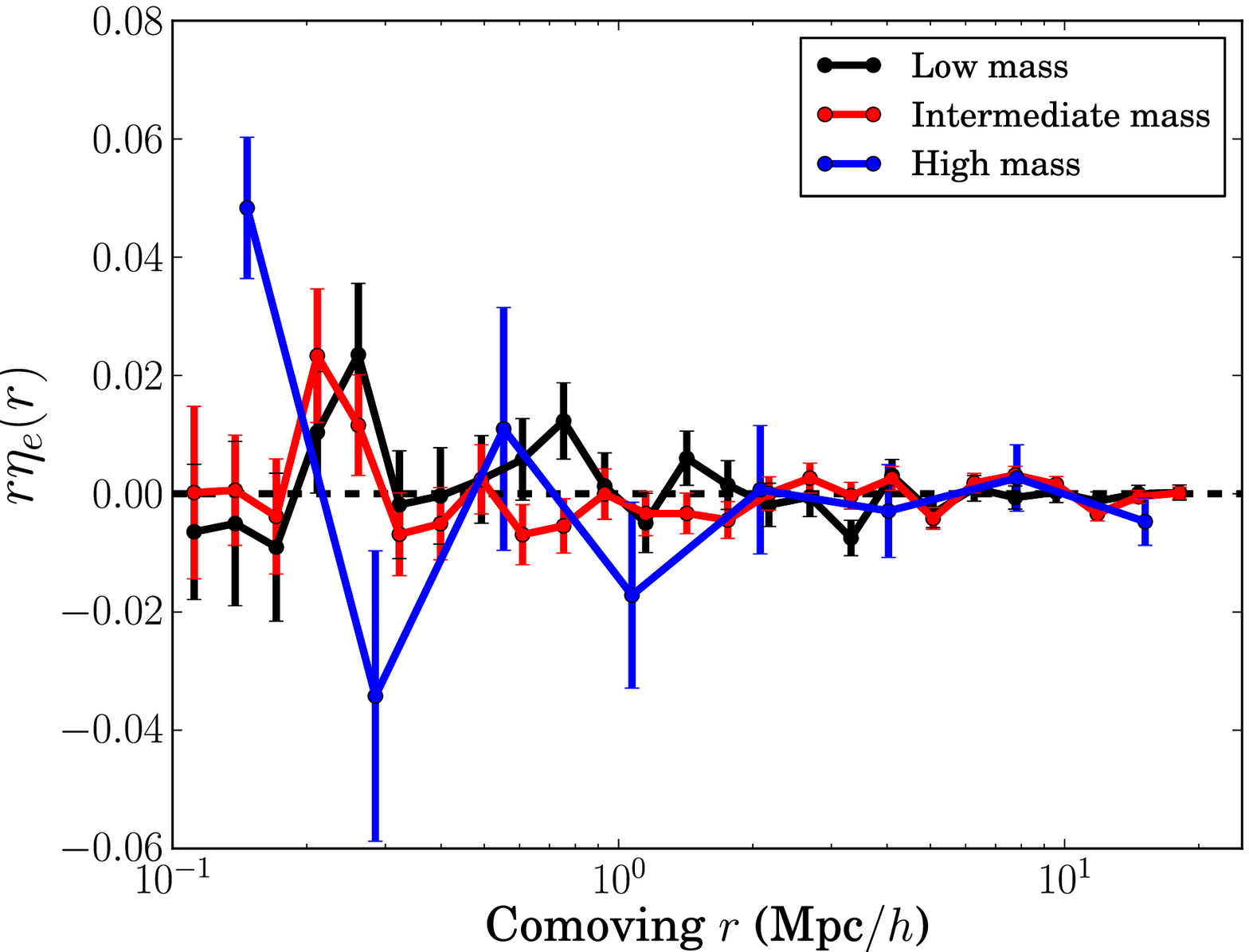}
\caption{Correlation between minor axis orientation and separation vector as a function of separation using randomized positions. Black lines correspond to low mass galaxies ($10^9\,{\rm M}_\odot<M_*<10^{9.5}\,{\rm M}_\odot$); red lines, to intermediate mass galaxies ($10^{9.5}\,{\rm M}_\odot<M_*<10^{10.6}\,{\rm M}_\odot$); and blue lines, to high mass galaxies ($M_*>10^{10.6}\,{\rm M}_\odot$).}
\label{fig:dominorrand}
\end{figure}

\begin{figure}
\centering
\includegraphics[width=0.47\textwidth]{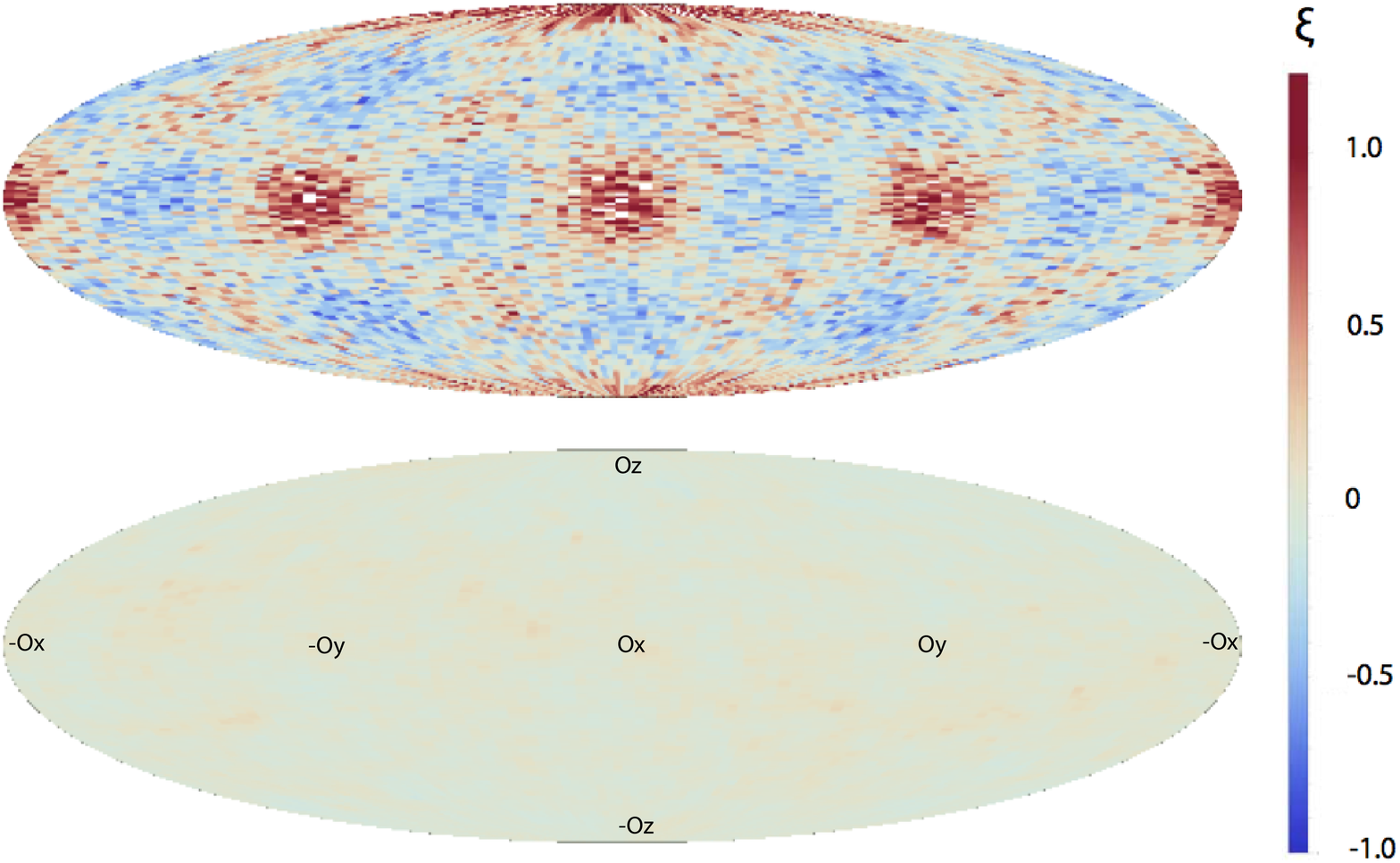}
\caption{Distribution of spins (top panel) and separation vectors of galaxy pairs separated by less than 25 Mpc$/h$ (bottom panel) on the sphere. The excess probability $\xi$ defined so that the PDF reads $P(\cos\theta,\phi)=(1+\xi)/4\pi$ is colour-coded from dark blue (-1) to dark red (+1). If the spins are clearly correlated with the box axes, this is not the case of the separation vectors for which the magnitude of the excess probability is smaller (lighter colours) and the pattern does not seem to be correlated with the grid.}
\label{fig:GL}
\end{figure}

Grid-locking creates correlations of the spins and shapes of galaxies with $\{x,y,z\}$. However, due to the periodic boundary conditions of the box, we expect that the grid-locking will average to null around any arbitrary point within the box. We test this hypothesis by randomizing the position of galaxies while computing the relative orientation of their spins and shapes around these random positions. Figure~\ref{fig:dominorrand} shows an example correlation for shapes defined from the orientation of the minor axis and galaxies in three mass bins. We do not find significant correlations when positions are randomized. Notice that this test is sufficient only under the assumption that the large-scale structure is not itself locked to the grid. This assumption was shown to be valid for the dark matter large-scale structure in \hagn \citep{dubois14}. We further test this assumption for galaxies in \hagn in Figure~\ref{fig:GL}, where we show the excess probability of spin alignments (top panel) and separation vectors alignments (bottom panel) with the grid for pairs of galaxies separated by less than $25$ Mpc$/h$ in the simulation. Spins are clearly correlated with the box axes, as demonstrated in Fig.~\ref{fig:box}, while this is not the case of the separation vectors.

\begin{figure}
\centering
\includegraphics[width=0.4\textwidth]{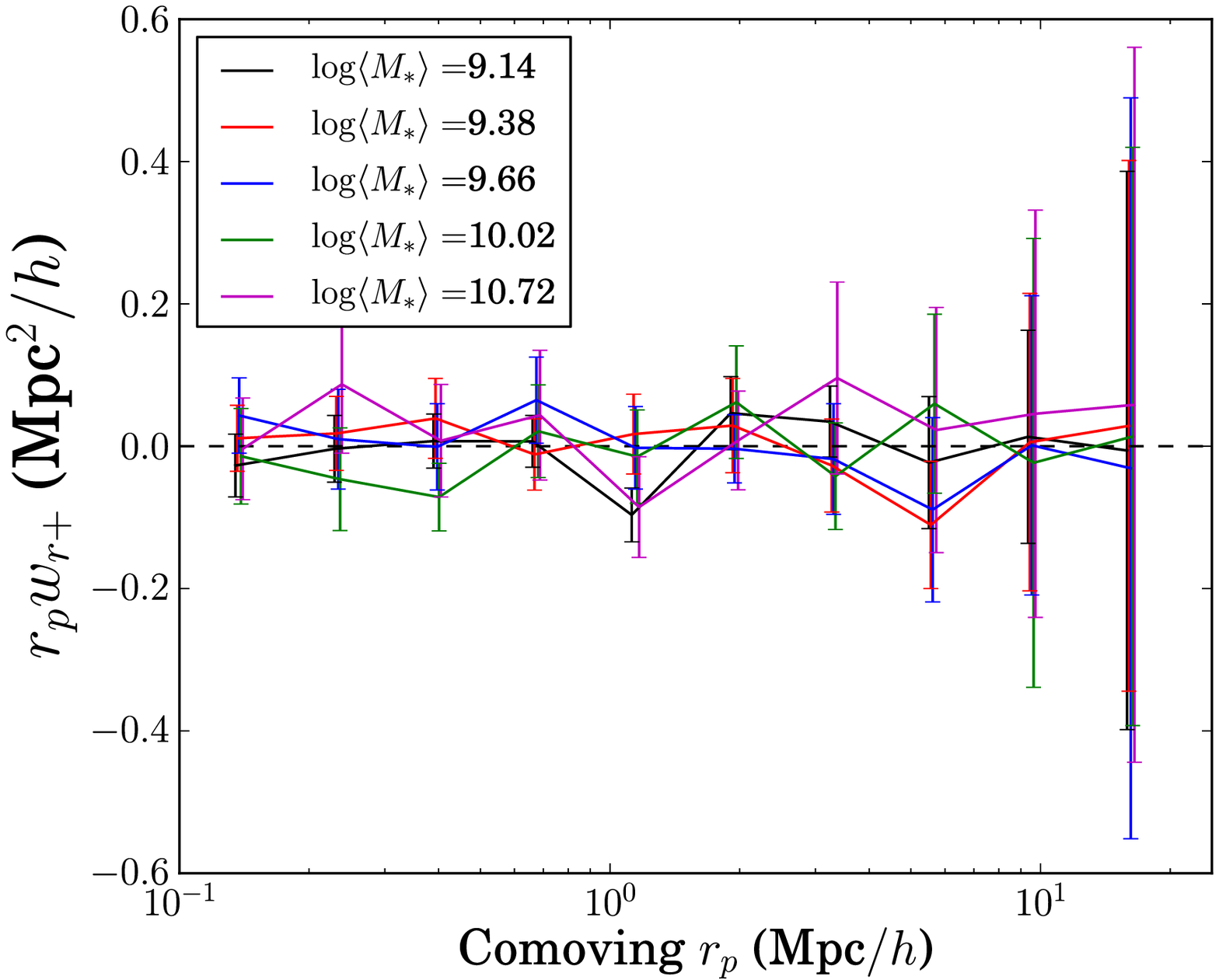}
\caption{Projected correlation function between random positions and $+$ component of the ellipticity. This result is consistent with null, which suggests that any systematics coming from grid-locking is not affecting our $\delta+$ or $\delta\times$ measurements. The different colours represent different bins in stellar mass and the points are slightly displaced to higher $r_p$ for each mass bin for visual clarity. The mean stellar mass in each bin is indicated in the legend.}
\label{fig:wprojrand}
\end{figure}

We apply a similar procedure to confirm that grid-locking does not contribute to $w_{\delta+}$. We obtain the projected correlation function of the random positions and the $+$ ($w_{r+}$) and $\times$ components of the shapes. Both of these correlations are consistent with null. This confirms that spherical averages remove the effect of grid-locking on position-shape correlations. Figure~\ref{fig:wprojrand} shows $w_{r+}$ for five mass bins as a function of $r_p$.

Notice that this procedure is not applicable to two-point auto-correlations of the spin or the minor axes. In other words, the impact of grid-locking on auto-correlations cannot be fully quantified by randomizing galaxy positions. This procedure would only measure the impact of uniform correlations of galaxy shapes or spins across $\{x,y,z\}$, but two galaxies clustered together could be increasingly affected by grid-locking. We present auto-correlations in Appendix~\ref{sec:auto}, along with possible interpretations of the signals as a result of physical alignments or grid-locking.


\section{Auto-correlations}
\label{sec:auto}

We have so far shown that the spins and shapes of galaxies tend to align with the grid, and that these grid-locking effects do not affect our measurements of position-shape (or position-spin) correlations. Unfortunately, the symmetry arguments that apply to position-shape correlations no longer hold when auto-correlations are considered. As a consequence, it is harder to separate the physical alignment signal and the grid-locking from auto-correlation of galaxies spins or shapes. In this section, we present those auto-correlations and we give the two possible interpretations of the signal.

The spin-spin (SS) correlation function measures the relative orientation of the angular momenta of two galaxies separated by a comoving distance vector ${\bf r}$,
\begin{equation}
\eta_S(r) = \langle |\hat{\bf s}({\bf x})\cdot \hat{\bf s}({\bf x}+{\bf r})|^2\rangle - {1}/{3}\,,
\label{eq:etaSS}
\end{equation}
and the ellipticity-ellipticity (EE) correlation function is given by 
\begin{equation}
\eta_E(r) = \langle |\hat{\bf e}({\bf x})\cdot \hat{\bf e}({\bf x}+{\bf r})|^2\rangle - {1}/{3}\,,
\label{eq:etaEE}
\end{equation}
where $\hat{\bf e}$ is the direction of the minor axis. Similarly to $\xi_{g+}$, defined in Equation~(\ref{eq:xigp}), we can define the auto-correlation function of $+$ components of the ellipticity
\begin{equation}
\xi_{++}(r_p,\Pi) = \frac{S_+S_+}{RR}\,,
\end{equation}
and its projection along the line-of-sight, $w_{++}$; and analogously for $\xi_{\times\times}$ and $w_{\times\times}$.

\begin{figure*} 
\centering
\includegraphics[width=0.33\textwidth]{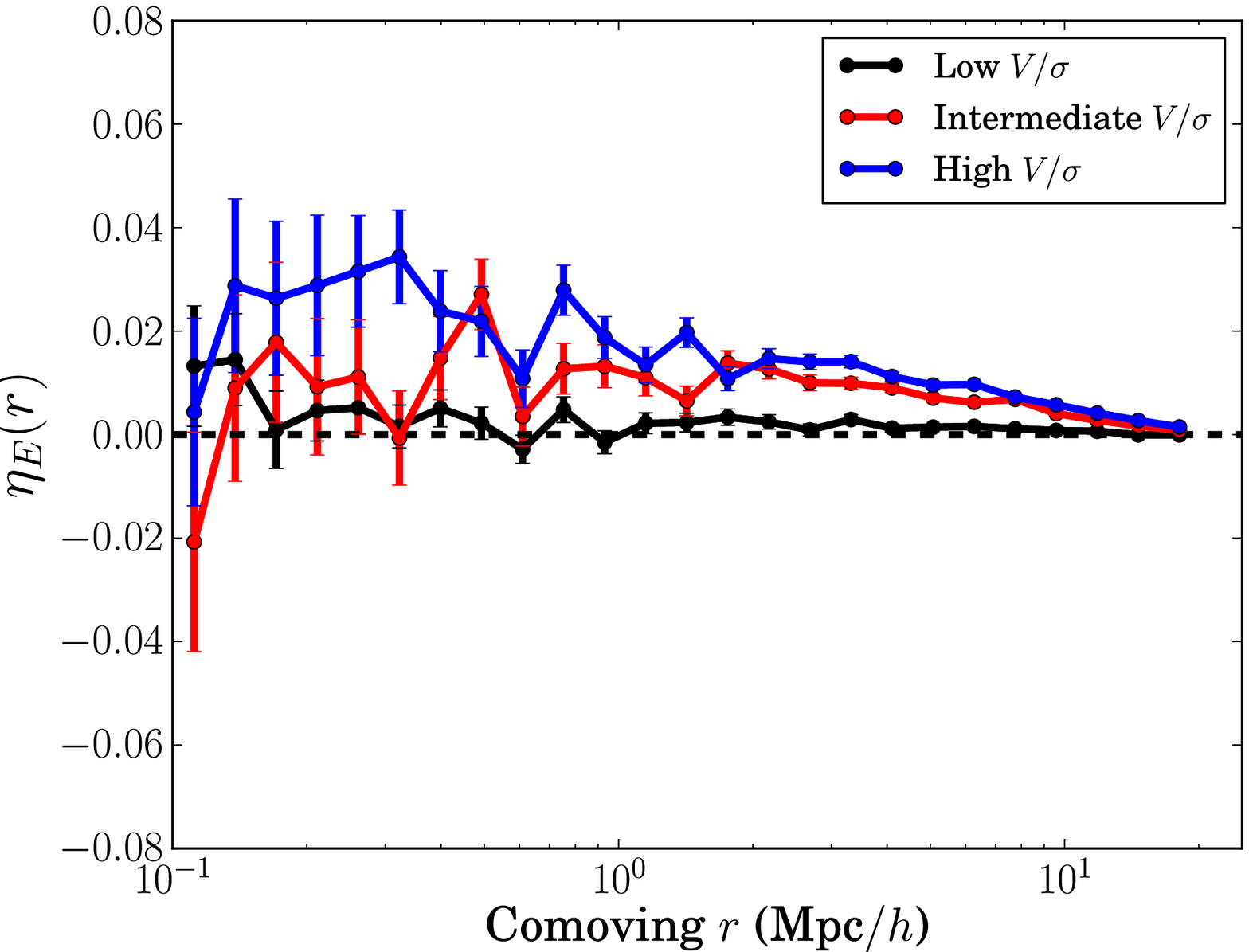}
\includegraphics[width=0.33\textwidth]{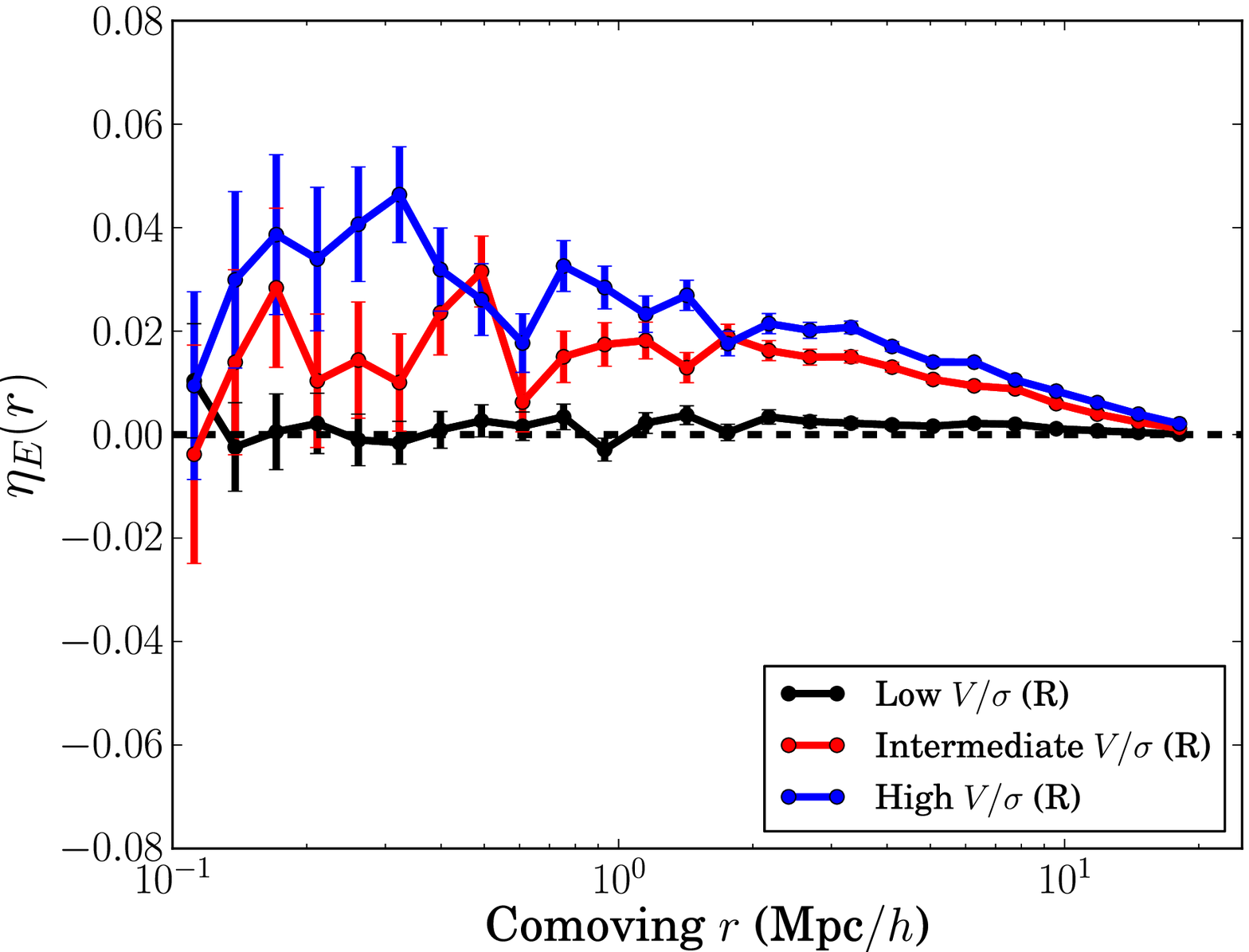}
\includegraphics[width=0.33\textwidth]{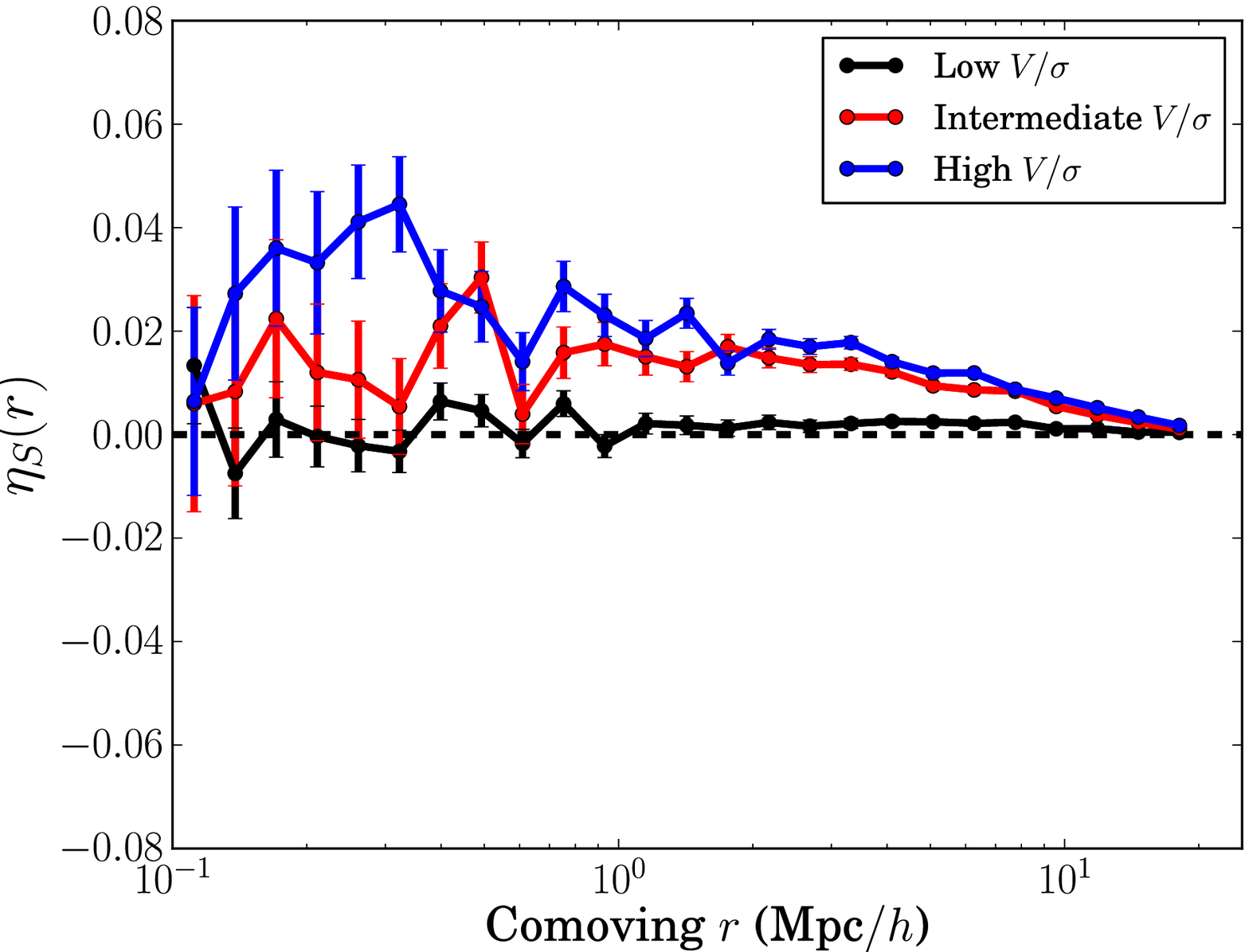}
\caption{Correlations between minor axes obtained from the simple inertia tensor (left panel), from the reduced inertia tensor (middle panel) and between spins (right panel) as a function of comoving separation. Black lines correspond to low $V/\sigma$ galaxies ($V/\sigma<0.55$); red lines, to intermediate $V/\sigma$ galaxies ($0.55<V/\sigma<0.79$); and blue lines, to high $V/\sigma$ galaxies ($V/\sigma>0.79$). }
\label{fig:spinstatvsig}
\end{figure*}

Figure~\ref{fig:spinstatvsig} shows the relative orientation of the spins of two galaxies (right panel) separated by a comoving distance of $r$. The correlation is significant for low and intermediate $V/\sigma$ galaxies. For pair separations $\lesssim10 \, h^{-1}\,\rm Mpc$, the spins tend to align with each other and the alignment decreases on larger scales. This trend is monotonic with $V/\sigma$. The results for the auto-correlation between minor axes or between spins are very similar when minor axes are computed using the reduced inertia tensor (middle panel of Figure~\ref{fig:spinstatvsig}). The signal slightly decreases when the simple inertia tensor is used (left panel). These three dimensional auto-correlations of galaxy orientations seem to be dominated by the contribution of disc-like galaxies. In Section~\ref{sec:align3d}, we saw that disc galaxies tend orient their spin (also reduced minor axis) tangentially around over-densities. Figure~\ref{fig:spinstatvsig} is qualitatively compatible with those results. Similarly, we find no auto-correlation between the shapes of low $V/\sigma$ galaxies in the left panel of Figure \ref{fig:spinstatvsig}.

\begin{figure*} 
\centering
\includegraphics[width=0.33\textwidth]{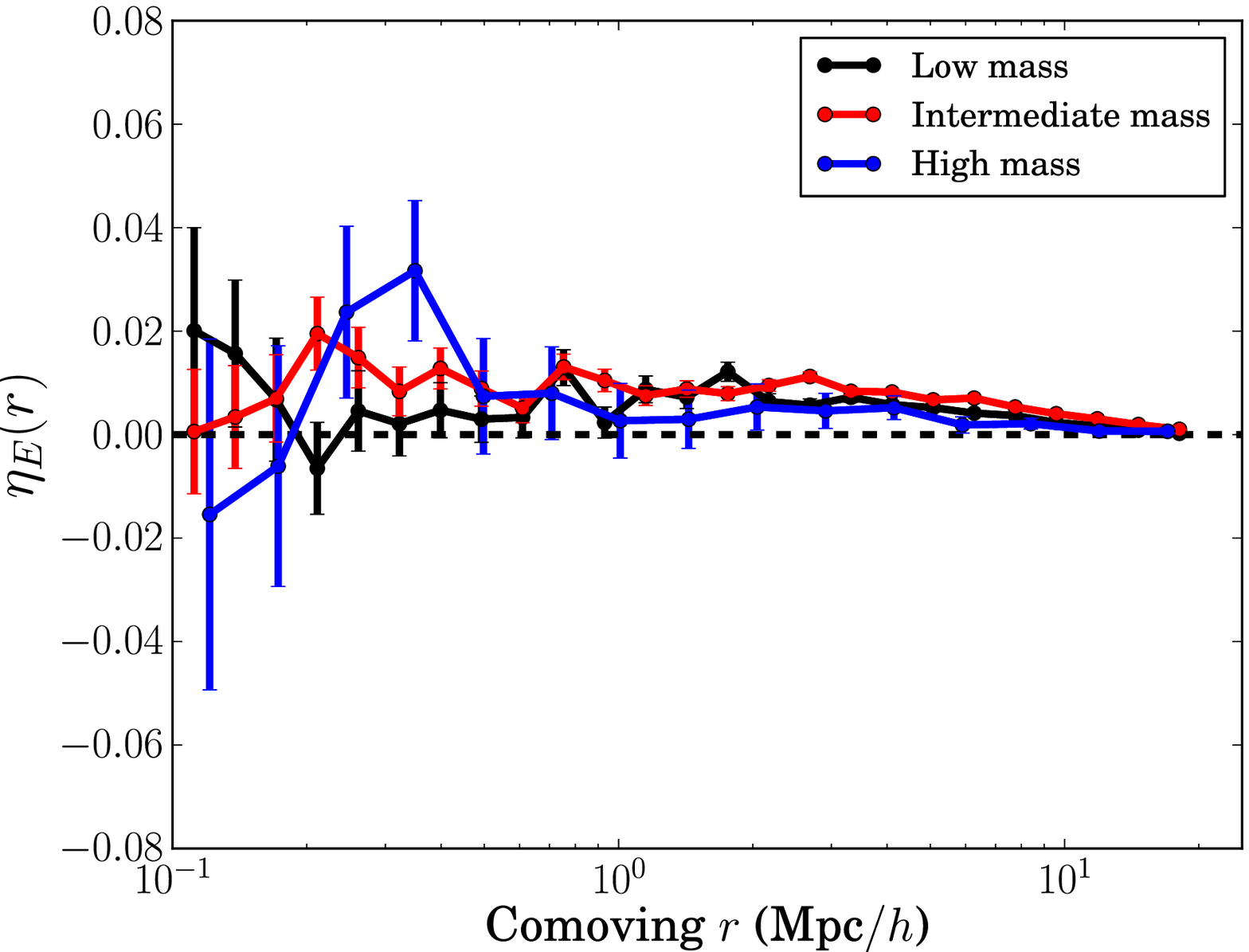}
\includegraphics[width=0.33\textwidth]{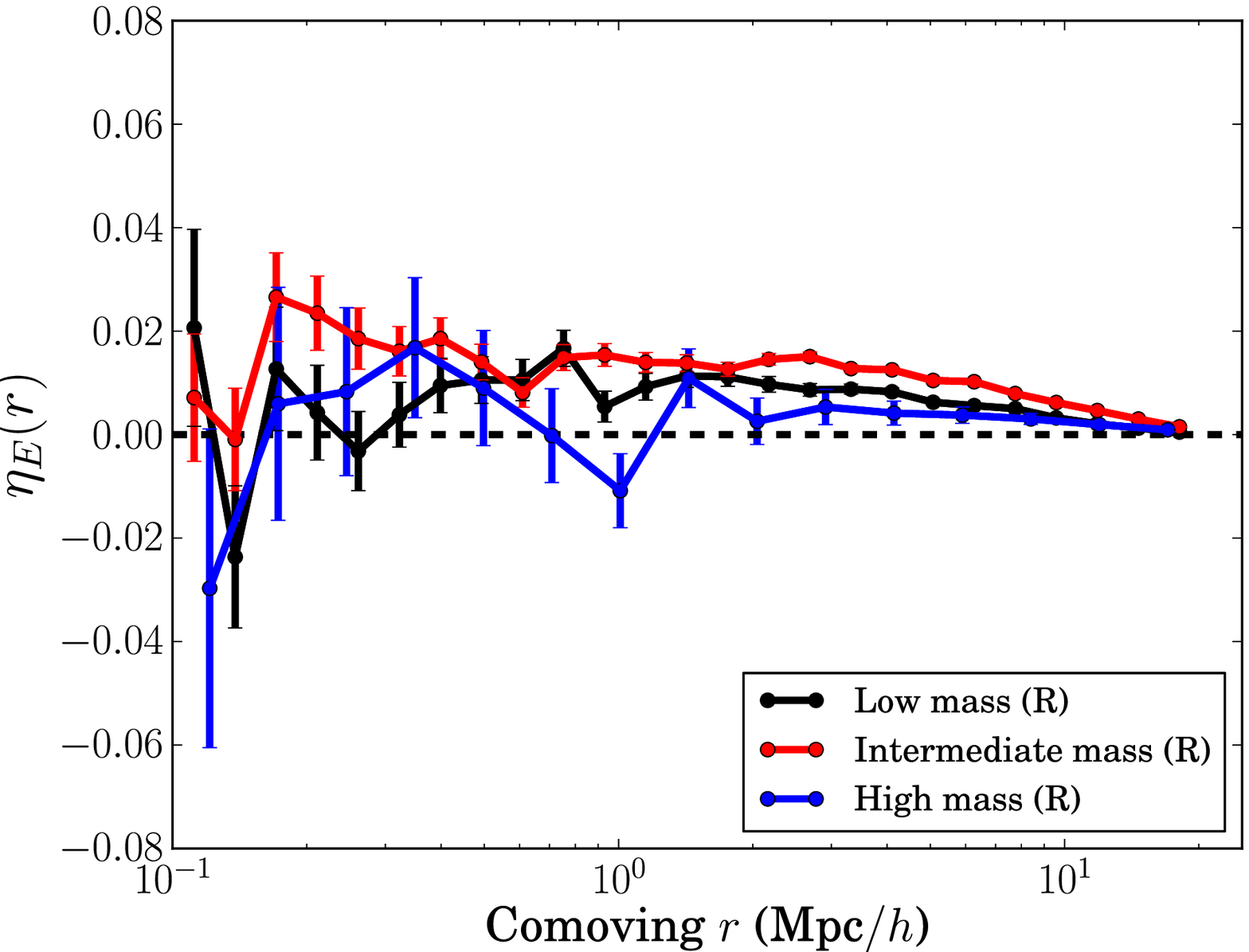}
\includegraphics[width=0.33\textwidth]{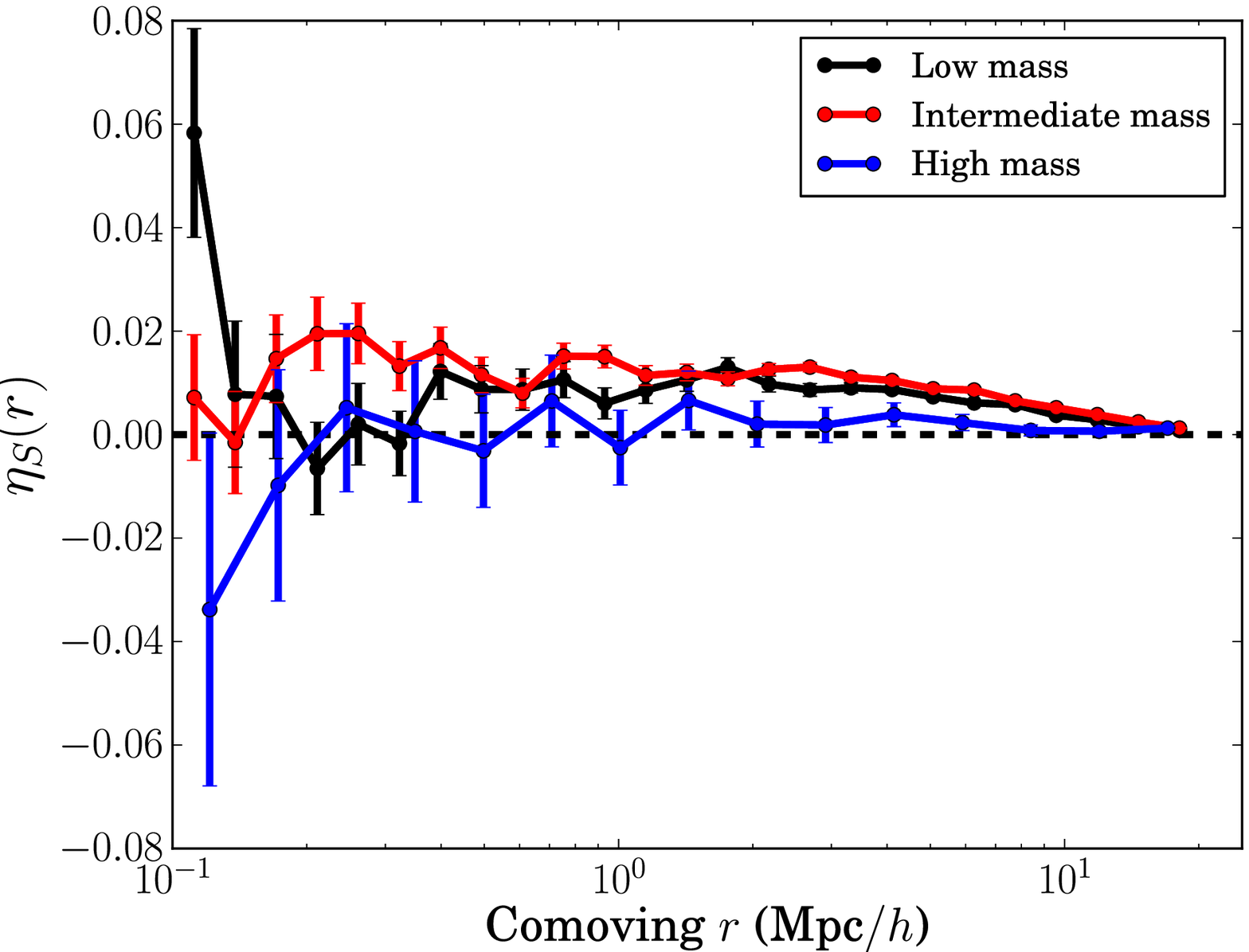}
\caption{Correlations between minor axes obtained from the simple inertia tensor (left panel), from the reduced inertia tensor (middle panel) and between spins (right panel) as a function of comoving separation. Black lines correspond to low mass galaxies ($10^9\,{\rm M}_\odot<M_*<10^{9.5}\,{\rm M}_\odot$); red lines, to intermediate mass galaxies ($10^{9.5}\,{\rm M}_\odot<M_*<10^{10.6}\,{\rm M}_\odot$); and blue lines, to high mass galaxies ($M_*>10^{10.6}\,{\rm M}_\odot$). }
\label{fig:spinstat}
\end{figure*}

We show the orientation-separation correlations as a function of mass in Figure~\ref{fig:spinstat}. Intermediate and low mass galaxies tend to have a stronger auto-correlation than high mass galaxies. The signal is stronger for intermediate mass galaxies, which is expected both from grid-locking and from the physical signal due to the contribution of the high $V/\sigma$ population. Similar qualitative results were obtained by \citet{Codis14} at $z=1.2$ using the \hagn simulation with the same mass selection (see their Figure 8 for a direct comparison). The trend with mass is similar, but the amplitude of the signal is higher in our case, suggesting that the spin auto-correlation increases at lower redshift and at fixed mass. \citet{Codis14} also found that separating the galaxy population by $u-r$ colour into red and blue galaxies resulted in a significant spin alignment signal for blue galaxies, while no spin alignment was found for red galaxies. This is also in agreement with our results, given the strong correlation between colour and $V/\sigma$. 
 

\begin{figure*} 
  \centering
    \includegraphics[width=0.33\textwidth]{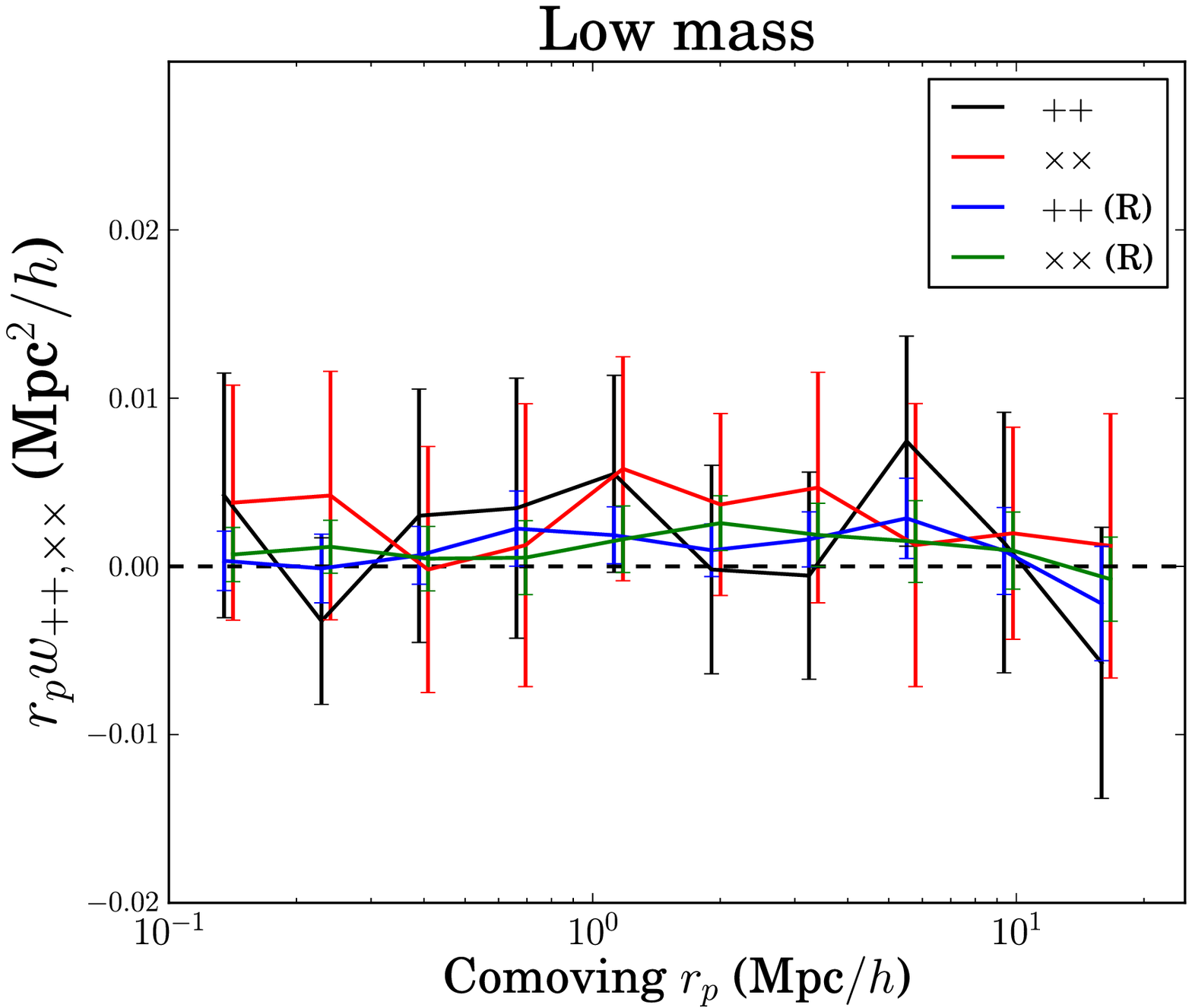}
    \includegraphics[width=0.33\textwidth]{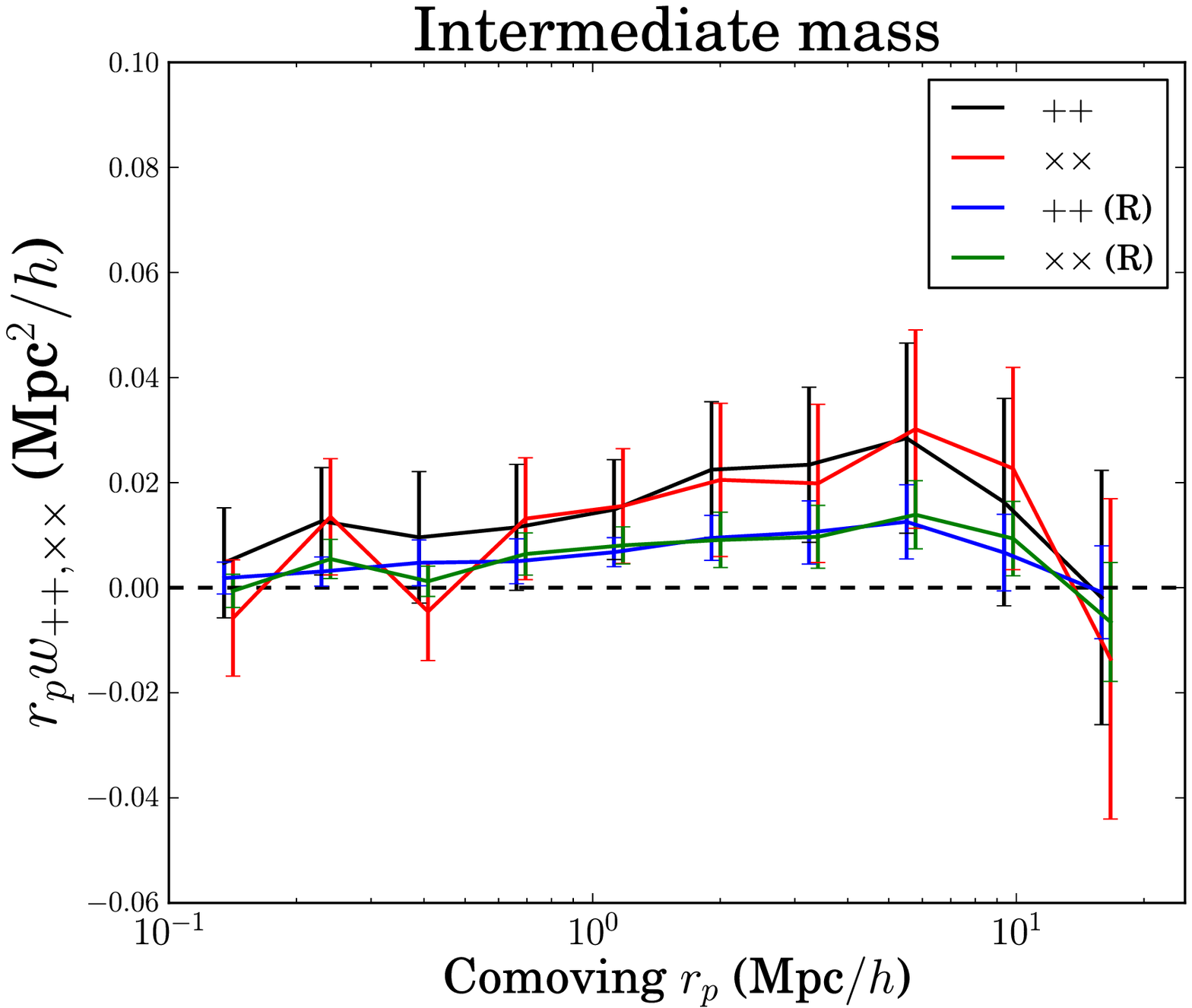}
    \includegraphics[width=0.33\textwidth]{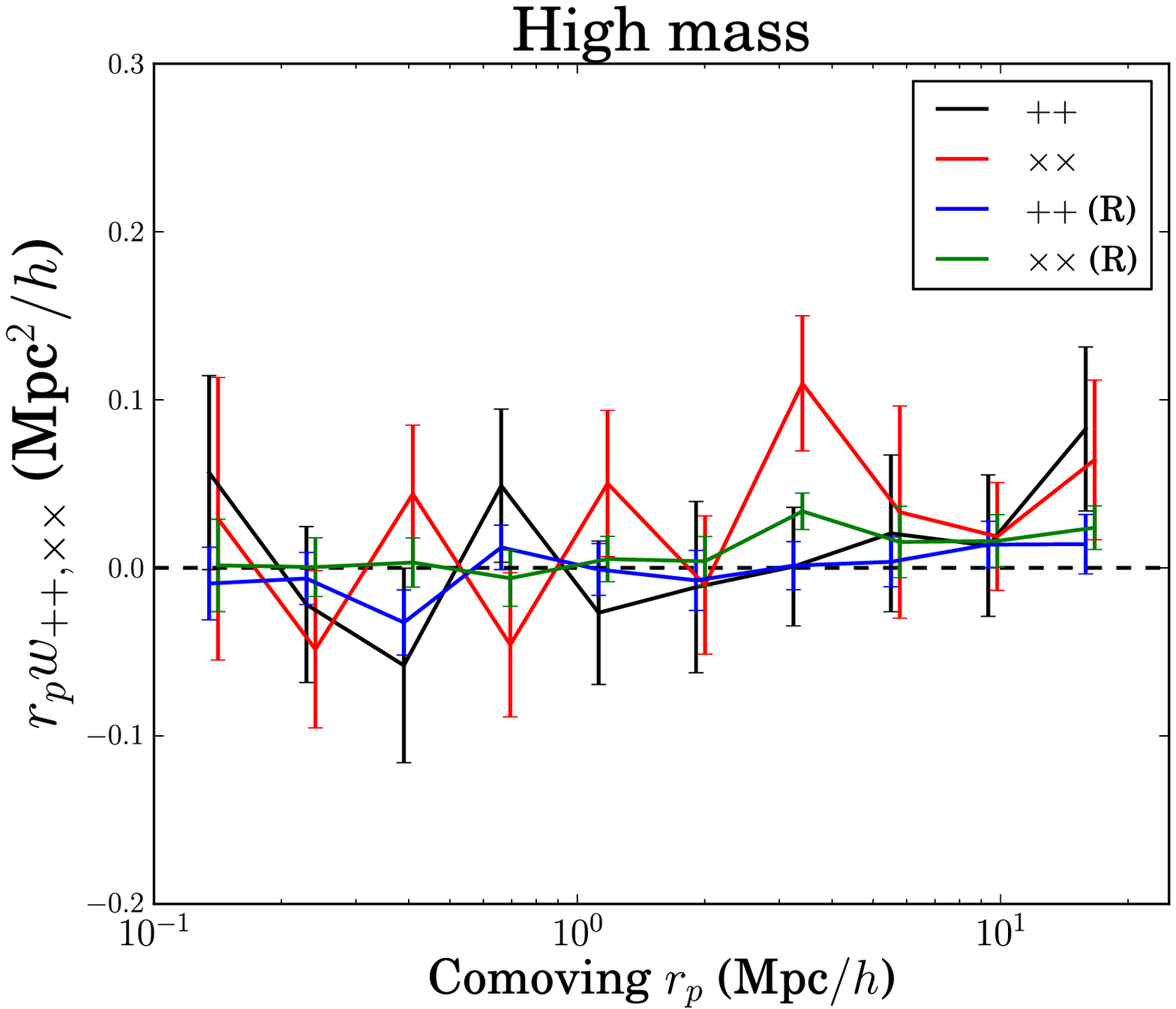}
    
    \includegraphics[width=0.33\textwidth]{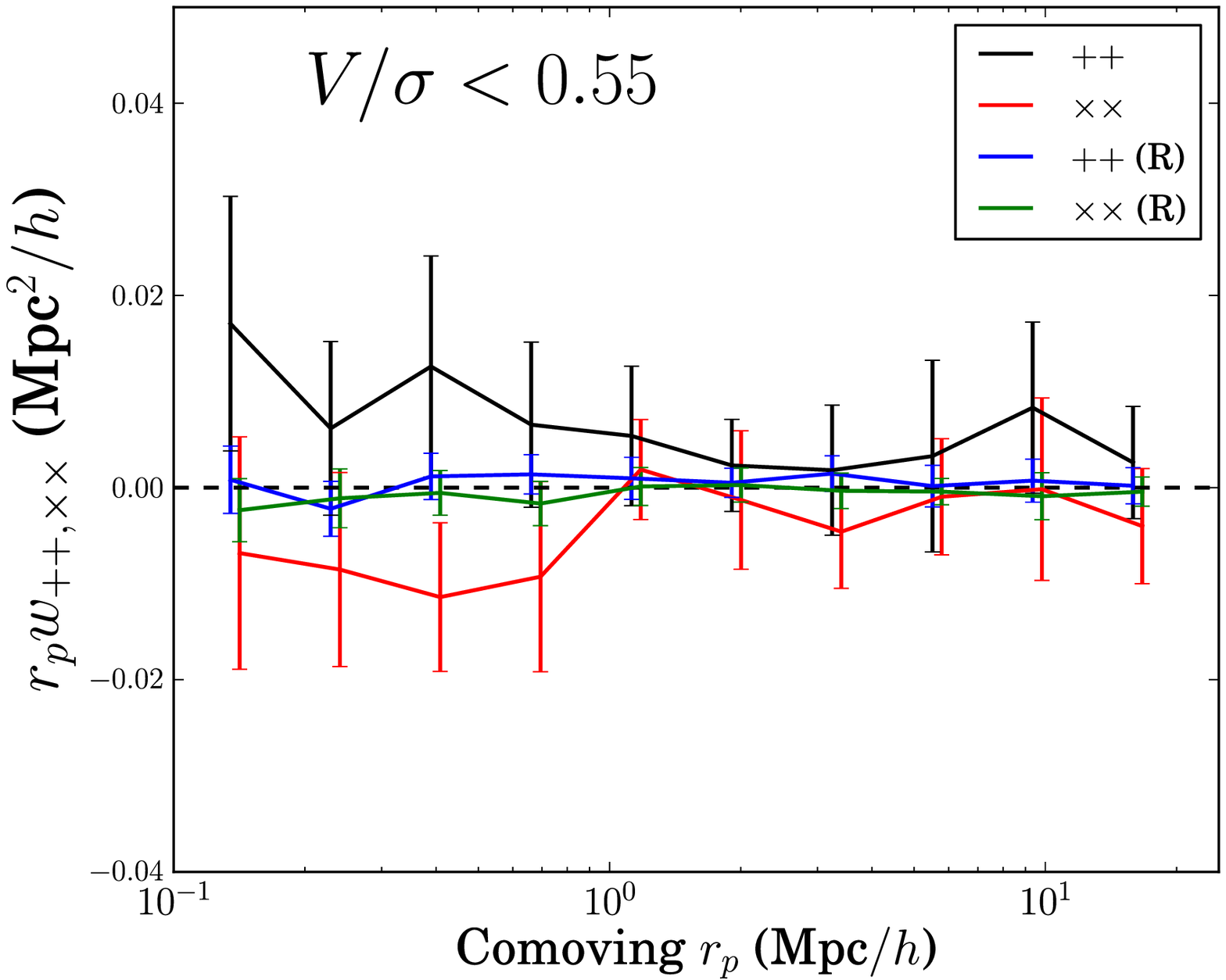}
    \includegraphics[width=0.33\textwidth]{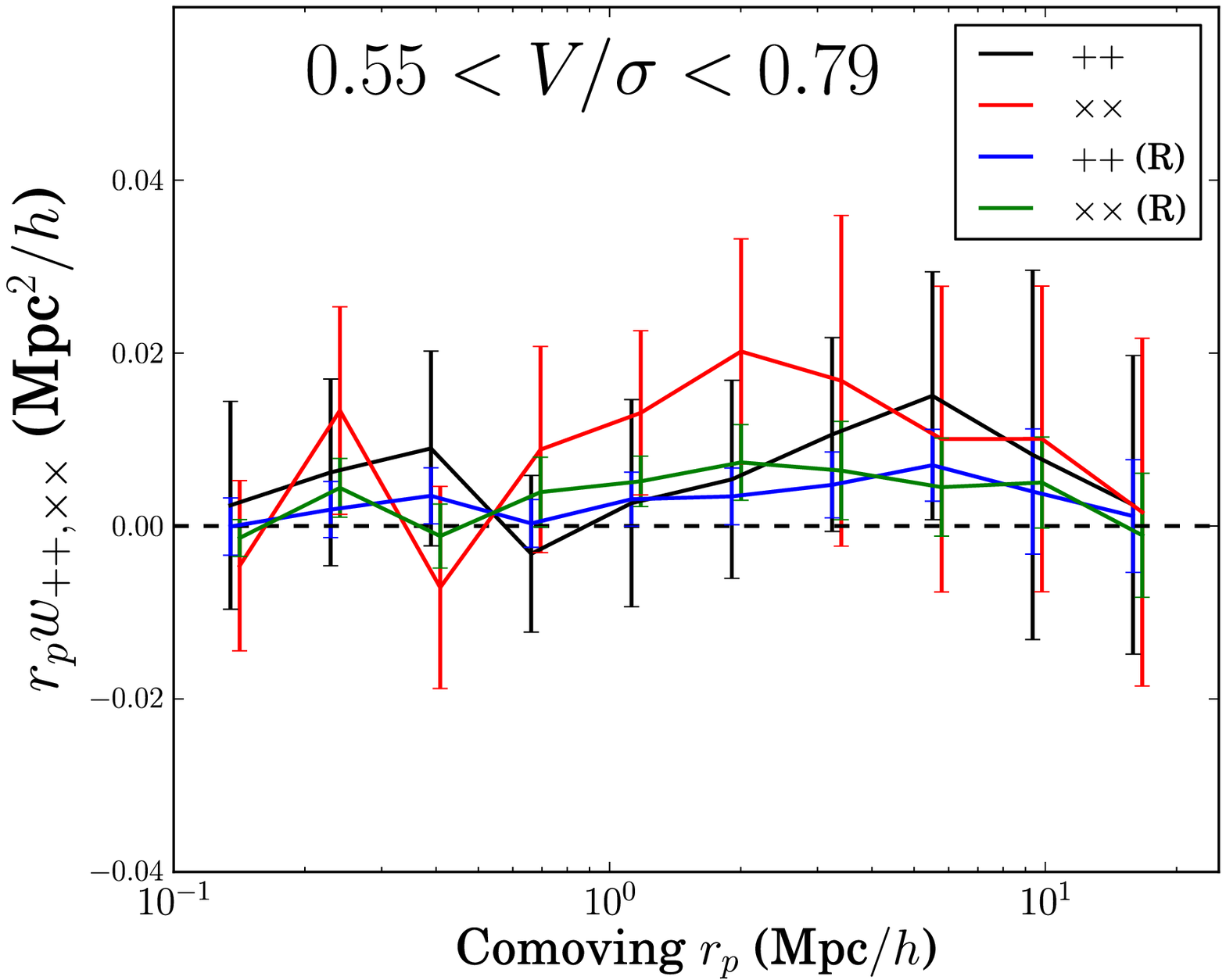}
    \includegraphics[width=0.33\textwidth]{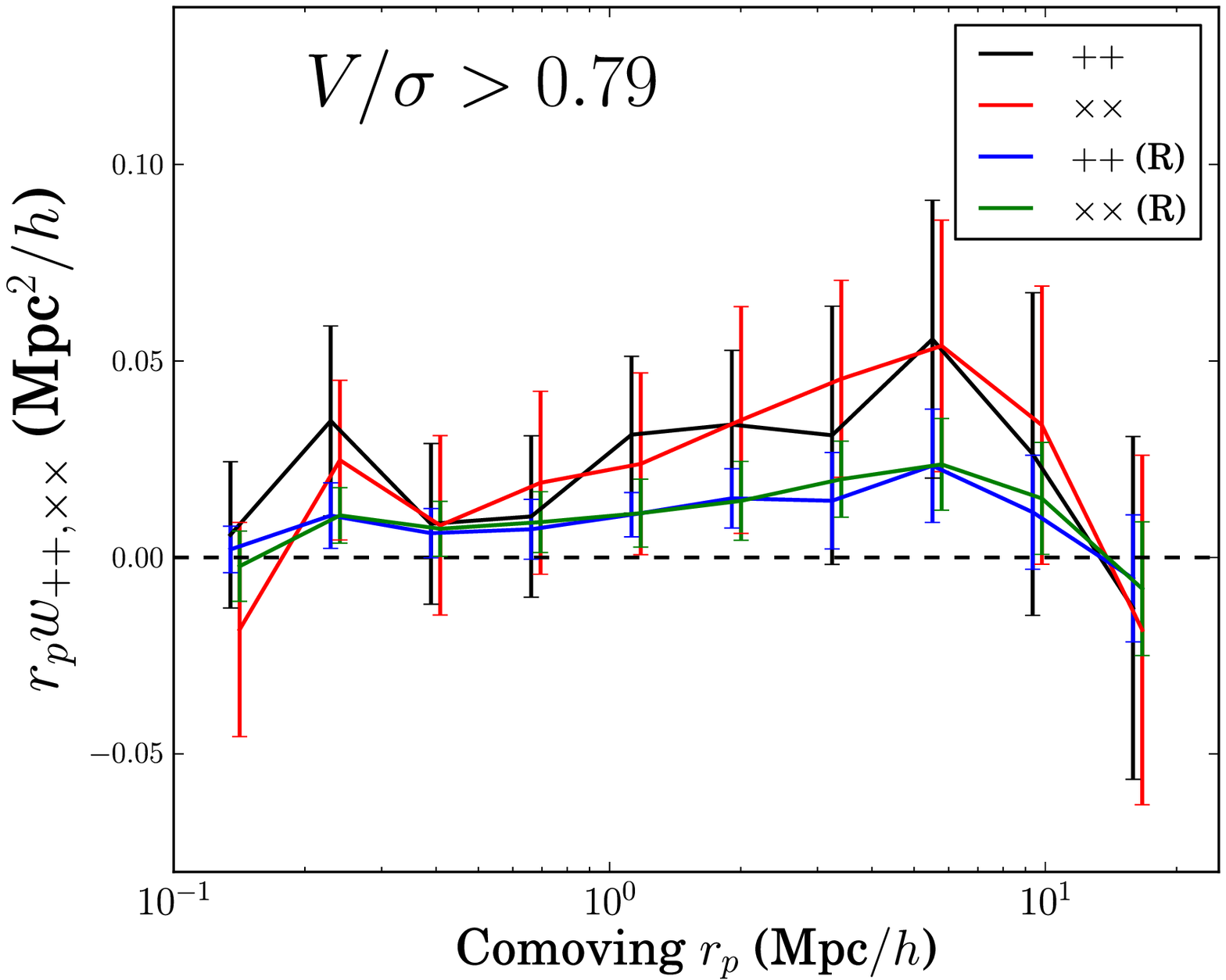}
  \caption{Projected auto-correlation functions of $+$ and $\times$ components of the shape as a function of projected radius in three mass bins (upper panels) and three $V/\sigma$ bins (lower panels). In all panels, the $++$ correlation is shown in black for shapes measured using the simple inertia tensor; correspondingly, the $\times\times$ correlation is shown in red. All panels show the impact of using the reduced inertia tensor (blue lines for $++$ and green lines for $\times\times$). All $\times\times$ correlations are arbitrarily displaced to $5\%$ higher $r_p$ for visual clarity. }
\label{fig:shapeshape}
\end{figure*}

The projected auto-correlation functions of galaxy shapes in three mass bins and three $V/\sigma$ are shown in Figure~\ref{fig:shapeshape} decomposed into $E$-modes ($++$) and $B$-modes ($\times\times$). Qualitatively, we observe that the projected auto-correlation of shapes is more significant for intermediate mass galaxies and high $V/\sigma$. This is consequence of a selection effect: galaxies with intermediate masses also tend to have higher $V/\sigma$ in Figure~\ref{fig:vsig}. We also find that the amplitude of the auto-correlation decreases when the reduced inertia tensor is used. We attribute this to the fact that $w_{++}$ and $w_{\times\times}$ carry a double weighting by galaxy ellipticities, which are also more round in the reduced case. 

Both $E$-modes and $B$-modes are present with similar amplitudes and scale-dependence in the top middle and right bottom panels of Figure~\ref{fig:shapeshape}. Using the diagonals of the jackknife covariance, we find that the signal is different from null at the $89\%$ C.L. ($85\%$ C.L.) for the $E$-modes ($B$-modes) of intermediate mass galaxies when using the simple inertia tensor; and higher ($99\%$ C.L.) using the reduced inertia tensor. For high $V/\sigma$ galaxies, the signal is different from null at the $79\%$ C.L. for $E$-modes and $81\%$ for $B$-modes using the simple inertia tensor; and higher ($>93\%$ C.L.) using the reduced inertia tensor. All other bins of mass and $V/\sigma$ have auto-correlations generally consistent with being null at the $2\sigma$ level.

Overall, we find that auto-correlations of galaxy shapes or spins are dominated by the contribution of disc-like galaxies. We find no contribution from spheroidals to these auto-correlations. Moreover, we find equal contribution of $E$-modes and $B$-modes in the projected correlation functions. If the signal is to be interpreted as a physical signal, the presence of an auto-correlation of disc-like tracers is expected, and the trend with mass/dynamics agrees with that found in Section~\ref{sec:align3d}. 

On the other hand, the absence of an auto-correlation for the shapes of spheroidals seems inconsistent with the presence of auto-correlations for discs. This could suggest that at least part of this signal is produced by grid-locking. $E$-mode and $B$-mode auto-correlations could also be produced by both a physical signal or by grid-locking. In the latter case, $B$-modes would be a consequence of preferential alignments of the simulated galaxies with the diagonals of the grid. Ideally, two-point correlations such as those presented in Figures \ref{fig:spinstatvsig} and \ref{fig:shapeshape} could potentially be used to constrain a simulation-dependent model of grid-locking. However, there is not sufficient evidence that current intrinsic alignment models predict the correct relation between density-shape and shape-shape correlations, especially on such small scales as probed in this work. Hence separating the grid-locking from the physical alignment signal in auto-correlations remains a difficult task for adaptive-mesh-refinement codes like {\sc ramses}.

\end{document}